\documentclass[aps,twocolumn,prd,showpacs,showkeys,preprintnumbers,superscriptaddress,nobibnotes,floatfix,longbibliography,nofootinbib]{revtex4-2}
\pdfoutput=1
\usepackage{amsmath}
\usepackage{amsfonts}
\usepackage{amssymb}
\usepackage{mathrsfs}
\usepackage{graphicx}
\usepackage{color}
\usepackage{hyperref}
\usepackage{multirow}
\usepackage{slashed}
\usepackage{epsf}

\usepackage{graphicx}
\usepackage{caption}
\usepackage{subcaption}

\usepackage{color}

\hypersetup{
   colorlinks=true,       
   linkcolor=blue,        
   citecolor=red,         
   filecolor=magenta,      
   }

\def\to{\rightarrow}

\begin{document}

\title{
Revisiting the Realistic Intersecting D6-Brane Model with  positive and negative $\mu$ Terms}

\author{Imtiaz Khan}
\email{ikhanphys1993@gmail.com}
\affiliation{Department of Physics, Zhejiang Normal University, Jinhua, Zhejiang 321004, China}
\affiliation{Zhejiang Institute of Photoelectronics, Jinhua, Zhejiang 321004, China}

\author{Ali Muhammad}
\email{alimuhammad@phys.qau.edu.pk}
\affiliation{CAS Key Laboratory of Theoretical Physics, Institute of Theoretical Physics, Chinese Academy of Sciences, Beijing 100190, China}
\affiliation{School of Physical Sciences, University of Chinese Academy of Sciences, No. 19A Yuquan Road, Beijing 100049, China}

\author{Tianjun Li}
\email{tli@mail.itp.ac.cn}
\affiliation{CAS Key Laboratory of Theoretical Physics, Institute of Theoretical Physics, Chinese Academy of Sciences, Beijing 100190, China}
\affiliation{School of Physical Sciences, University of Chinese Academy of Sciences, No. 19A Yuquan Road, Beijing 100049, China}
\affiliation{School of Physics, Henan Normal University, Xinxiang 453007, P. R. China}

\author{Shabbar Raza}
\email{shabbar.raza@fuuast.edu.pk}
\affiliation{Department of Physics, Federal Urdu University of Arts, Science and Technology, Karachi 75300, Pakistan}

\begin{abstract}
In light of current constraints from supersymmetry (SUSY) searches within the LHC as well as findings from direct dark matter detection experiments such as LUX-ZEPLIN (LZ), we revisit the three-family Pati–Salam model derived from intersecting D6-branes in Type IIA string theory compactified on the \( T^6/(\mathbb{Z}_2 \times \mathbb{Z}_2) \) orientifold, known for its realistic low-energy phenomenology. {Because the muon anomalous magnetic moment might be in accordance with the Standard Model prediction, }
we conduct a comprehensive scan over the model’s parameter space for each sign of the Higgsino mass parameter, \( \mu < 0 \) and \( \mu > 0 \). We found that a gravitino mass is typically greater than 1.5~TeV in both scenarios while simultaneously satisfying the LHC SUSY bounds, B-physics observables, along with the Higgs mass constraint. Within the experimentally viable region of the parameter space, the mass spectra of sparticles are found to fall within the following ranges: Gluinos lie in the range of 2--18~TeV, the first- and second-generation squarks and sleptons span 3--16~TeV and 1--6~TeV, respectively. For third-generation sfermions, the lightest stop, which can satisfy the dark matter relic density through neutralino-stop coannihilation consistent with the {Planck} 5$\sigma$ bounds, has a mass in the range of 0.5--1.2~TeV. Note that the lightest neutralino will be as heavy as 2.9~TeV. Additionally, the lightest stau could be as light as 200~GeV, as well as heavy up to 5.2~TeV. We further identify several viable mechanisms, including multiple coannihilation channels and resonance mechanisms, and the observed dark matter relic abundance is successfully realized.
\end{abstract}
\maketitle

	\section{Introduction}
Supersymmetry (SUSY) offers a compelling resolution for the gauge hierarchy  problem inherent in the Standard Model (SM). Supersymmetric extensions of the SM (SSMs), particularly those preserving $R$-parity, address several outstanding issues in particle physics. Notably, they facilitate gauge coupling unification~\cite{gaugeunification}, and enable radiative electroweak symmetry breaking, primarily caused due to the large top quark Yukawa coupling. Beyond the hierarchy problem, one of the most significant motivations for new physics occurring on the TeV scale is the search for insight into the nature of dark matter (DM), specifically, a thermal freeze-out mechanism during the early Universe. In $R$-parity-conserving SSMs, the Lightest Supersymmetric Particle (LSP)—typically the lightest neutralino, the gravitino, or similar—is stable and thus a natural DM candidate~\cite{Goldberg:1983nd}. Although these candidates must satisfy stringent constraints imposed by direct detection (DD) experiments and other DM searches.

Another foundational motivation for this study is the pursuit of a consistent theory of quantum gravity, where string theory emerges as a leading framework. String phenomenology aims to construct realistic low-energy theories, such as the SM or SSMs, from string theory with stabilized moduli, absence of chiral exotics, and predictive features testable at colliders like the LHC. This work focuses on DM phenomenology in the framework of the intersecting D-brane model~\cite{Berkooz:1996km, Ibanez:2001nd, Blumenhagen:2001te, CSU, Cvetic:2002pj, Cvetic:2004ui, Cvetic:2004nk, Cvetic:2005bn, Chen:2005ab, Chen:2005mj, Blumenhagen:2005mu}. In these constructions, realistic fermion Yukawa couplings are most naturally occur in the Pati-Salam gauge group~\cite{PS}. Specifically, three-family Pati-Salam models have been systematically built in Type IIA string theory on the $\mathbf{T^6/(\mathbb{Z}_2 \times \mathbb{Z}_2)}$ orientifold using intersecting D6-branes~\cite{Cvetic:2004ui}. Among these, a phenomenologically viable model has been identified that exhibits tree-level gauge coupling unification near the string scale and where the Pati-Salam symmetry is successfully broken down to the SM. The model also accommodates realistic fermion masses and mixing, while excess chiral exotics can be decoupled via the Higgs mechanism or strong dynamics. Furthermore, the resulting sparticle spectrum is potentially accessible at the  Large Hadron Collider (LHC), and the correct relic abundance of DM can be achieved through a neutralino LSP~\cite{Chen:2007px, Chen:2007zu, Li:2014xqa}. In summary, this construction stands out as one of the most promising globally consistent string-derived models. It offers a coherent phenomenological framework from the string scale to the electroweak scale, incorporating the Minimal Supersymmetric Standard Model (MSSM) spectrum and providing testable predictions at current and future experiments. 
       
{In light of the possible agreement between the experimental measurement of the muon's anomalous magnetic moment and the SM prediction, we carry out an extensive scan of the model’s parameter space, systematically analyzing scenarios with both signs of the Higgsino mass parameter, \( \mu > 0 \) and \( \mu < 0 \).}
The non-universality of gaugino masses, when considered in conjunction with the sign of the higgsino mass parameter\(\mu\), opens up a rich landscape for phenomenological exploration within the framework of supersymmetric \(SU(4)_c \times SU(2)_L \times SU(2)_R\) (4-2-2) models. The sign of \(\mu\) plays a pivotal role in determining various low-energy observables and has deep implications for the viability of specific SUSY spectra. In particular, Ref.~\cite{Gogoladze:2010fu} demonstrated how essential it is to select \(\mu < 0\) in order to obtain the proper threshold corrections for the bottom quark Yukawa coupling, which is crucial for obtaining viable third-generation Yukawa unification (t-b-\(\tau\) YU). A negative \(\mu\) facilitates a better match with the low-energy bottom quark mass while maintaining a light SUSY spectrum compatible with unification constraints. {Moreover, the sign \(\mu\) has notable consequences for the anomalous magnetic moment of the muon \((g-2)_{\mu}\), as highlighted in Ref. \cite{Ahmed:2021htr}. Regarding supersymmetric theories, a positive \(\mu\) generally leads to constructive contributions to \((g-2)_{\mu}\), helping to explain the long-standing deviation observed between the SM prediction and experimental measurements. Assuming the realization of SUSY at the electroweak scale, the dominant contributions to the muon anomaly arise from loop diagrams involving smuon-neutralino and sneutrino-chargino exchanges. These loop-induced corrections can be approximately expressed as~\cite{KhalilS2017}:
\begin{equation}
\Delta a_\mu^{\text{SUSY}} \sim \frac{M_i \, \mu \, \tan\beta}{m_{\text{SUSY}}^4},
\label{eq:g-2}
\end{equation}
where \( M_i \) (\( i = 1,2 \)) are the electroweak gaugino masses, \( \mu \) is the Higgsino mass parameter, \( \tan\beta = \langle H_u \rangle / \langle H_d \rangle \) denotes the ratio of the vacuum expectation values of the two Higgs doublets, and \( m_{\text{SUSY}} \) represents the characteristic mass scale of SUSY particles circulating in the loops.
As seen from Eq.~\ref{eq:g-2}, the SUSY contribution to \( a_\mu \) is directly proportional to the sign and magnitude of \( \mu \). A positive \( \mu \) typically enhances this contribution, especially in models with light sleptons and electroweakinos, and is therefore consistent with previous reports of a discrepancy between SM predictions and experimental measurements~\cite{Muong-2:2023cdq, Ahmed:2021htr}. However, in scenarios where the muon anomaly is reconciled within the SM or becomes less statistically significant, the negative \( \mu \) regime regains phenomenological relevance.
Recently, the Muon \( (g{-}2) \) Experiment at Fermilab (E989) has reported an updated and highly precise measurement of the positive muon’s magnetic anomaly, based on data collected between 2020 and 2023~\cite{Muong-2:2023cdq, Muong-2:2021ojo, Muong-2:2025xyk}:
\[
a_\mu = 116\,592\,0710(162) \times 10^{-12} \quad \text{(139 ppb)}.
\]
When combined with earlier measurements, the refined estimate becomes~\cite{Muong-2:2025xyk}:
\[
a_\mu = 116\,592\,0705(148) \times 10^{-12} \quad \text{(127 ppb)}.
\]
The current world average, now largely determined by Fermilab results, is~\cite{Muong-2:2025xyk}:
\[
a_\mu^{\text{exp}} = 116\,592\,0715(145) \times 10^{-12} \quad \text{(124 ppb)}.
\]
This represents over a fourfold improvement in experimental precision relative to prior determinations.
In parallel, advancements in lattice QCD have significantly improved the theoretical prediction for the leading-order hadronic vacuum polarization (LO HVP) contribution. A consolidated lattice-QCD estimate now achieves a precision of approximately 0.9\%, yielding a refined Standard Model prediction~\cite{Aliberti:2025beg}:
\[
a_\mu^{\text{SM}} = 116\,592\,033(62) \times 10^{-11} \quad \text{(530 ppb)}.
\]
The updated difference between the experimental and theoretical values is~\cite{Muong-2:2025xyk, Aliberti:2025beg}:
\[
\Delta a_\mu = a_\mu^{\text{exp}} - a_\mu^{\text{SM}} = 38.5(63.673) \times 10^{-11},
\]
corresponding to a deviation of only \( 0.6\sigma \). This effectively resolves the previous 4.2\( \sigma \) tension, indicating consistency between the SM prediction and experimental observations at the current level of precision.
Therefore, the supersymmetry scenario with negative $\mu~ ( \mu < 0 )$ is interesting right now.
Additionally, the sign of \(\mu\) is critical for realizing the Higgs and Z-pole mediated DM annihilation channels, which are subject to stringent constraints from the LHC electroweakino searches and DD DM experiments such as LZ~\cite{Khan:2025azf,Barman:2022jdg}. All these conclude that the phenomenological studies of SUSY GUTs are timely and crucial for the $\mu<0$.} Furthermore, the soft supersymmetry (SUSY) breaking terms, as considered in \cite{Li:2014xqa} with a focus on natural SUSY and subsequently updated in \cite{Sabir:2022hko}, are critical to the phenomenological study of intersecting D6-brane models \cite{Cvetic:2004ui}, especially in the context of current LHC SUSY searches and DD DM constraints. Motivated by these developments, a comprehensive analysis of the phenomenology of intersecting D6-brane models, compactified on the $\mathbf{T^6/(\mathbb{Z}_2 \times \mathbb{Z}_2)}$ orientifold, becomes particularly relevant, with particular emphasis on the $\mu>0$ and $\mu<0$ scenarios. These constructions provide a natural setting for accommodating the required non-universal gaugino masses and offer a robust framework for addressing both collider and cosmological constraints through string-inspired model building. Moreover, as the LHC prepares for Run-3 and the commencement of a new data-taking period, a comprehensive reassessment of the current status of supersymmetric scenarios becomes increasingly timely and essential. Such an investigation is crucial for identifying the most promising and phenomenologically viable regions of the SUSY parameter space. These regions can serve as focal points for targeted searches at Run-3, thereby enhancing the discovery potential of the LHC and guiding future experimental strategies. In this study, we perform a comprehensive phenomenological analysis of the supersymmetric parameter space, focusing on both positive and negative values of the higgsino mass parameter \(\mu > 0\) and \(\mu < 0\). Our analysis incorporates constraints from current LHC SUSY searches and DD bounds on DM and employs the framework of modified soft supersymmetry-breaking terms as discussed in~\cite{Li:2014xqa}, with updated computational methodology based on~\cite{Sabir:2022hko}. The resulting viable parameter space satisfying all imposed experimental constraints, including the Higgs boson mass requirement, reveals interesting mass spectra for the first two generations of sparticles. Specifically, we observe gluino masses in the range \([2, 18]~\text{GeV}\), first-generation squark masses between \([3, 16]~\text{GeV}\), and slepton masses from \([1, 6]~\text{GeV}\).

 For the third generation sparricles spectrum, our analysis identifies several key mechanisms responsible for generating the correct DM relic density within the observed limits. These include:
\begin{itemize}
    \item \textbf{A/H-resonance annihilation channels}: Present in the \(\mu > 0\) scenario with the pseudoscalar Higgs mass \(m_{A/H} \approx 2~\text{TeV}\); notably absent in the \(\mu < 0\) case.
    \item \textbf{Chargino-neutralino coannihilation}: For \(\mu > 0\), the next-to-lightest supersymmetric particle (NLSP) chargino mass lies in the range \(0.7\)–\(2.5~\text{TeV}\); for \(\mu < 0\), it extends up to \(2.8~\text{TeV}\).
    \item \textbf{Stau-neutralino coannihilation}: The NLSP stau mass is found within \(0.2\)–\(1.8~\text{TeV}\) for \(\mu > 0\), and up to \(2.5~\text{TeV}\) for \(\mu < 0\).
    \item \textbf{Stop-neutralino coannihilation}: Viable solutions exist for stop masses between \(0.15\) and \(1.2~\text{TeV}\) in both \(\mu > 0\) and \(\mu < 0\) scenarios.
\end{itemize}

It is worth mentioning that the current LHC SUSY searches have already thoroughly probed a significant area of the parameter space, especially linked to the stop-neutralino coannihilation region. However, with the exception of a few chargino-neutralino solutions, the majority of the coannihilation scenarios are still compatible with the sensitivity of current and upcoming astrophysical dark matter detection investigations, including XENONnT, LZ current, and LZ 1000-days sensitivity, a future DD DM experiment  \cite{LZ:2022lsv,LZ:2018qzl, XENON:2023cxc}. 

	\section{A Realistic Pati-Salam Model of Intersecting D6-Branes on $\mathbf{T^6/(\mathbb{Z}_2\times \mathbb{Z}_2)}$ Orientifold}
	\label{model}
	
This work focuses on the Type IIA string theory's phenomenologically feasible Pati-Salam model, which is built from intersecting D6-branes ~\cite{Cvetic:2004ui}, incorporating revised soft SUSY-breaking (SSB) terms that were computed in ~\cite{Sabir:2022hko}. Neglecting CP-violating phases, the SSB terms induced by non-zero F-term associated with the dilaton ($F^S$) as well as the three complex structure moduli ($F^{U^i}$, where $i = 1,2,3$) can be parametrized using the angles $\Theta_1$, $\Theta_2$, $\Theta_3$, and $\Theta_4$, along with the gravitino mass $m_{3/2}$. Following, $\Theta_4 \equiv \Theta_S$ corresponds specifically to the dilaton contribution. The $\Theta_i$ parameters are subject to the normalization condition given in Ref~\cite{Li:2014xqa}, 
    \begin{widetext}
\begin{align}
	\sum_{i=1}^4\Theta^2_i=1.
	\label{Theta4}
\end{align}
 Based on these parameters, the set of SSB terms at the grand unification (GUT) scale are able to expressed as~\cite{Sabir:2022hko} 
\begin{align}
A_0 &= m_{3/2}(-0.292797 \Theta_1-1.43925 \Theta_2-0.573228 \Theta_3 +0.573228 \Theta_4) ,\nonumber\\ \widetilde{m}_{H_u}^2 & =\widetilde{m}_{H_d}^2=m_{3/2}^2(1.0-(1.5 \Theta_{3}^2)-(1.5 \Theta_{4}^2)),\nonumber\\
 M_1 & =m_{3/2} (0.519615 \Theta _1+0.34641 \Theta _2+0.866025 \Theta _3) ,\nonumber\\
M_2 & =m_{3/2} (0.866025 \Theta _2-0.866025 \Theta _4) ,\nonumber\\
M_3 & =m_{3/2} ( 0.866025 \Theta _2+0.866025 \Theta _3) ,\nonumber\\
 \widetilde{m}^2_{R}& = m_{3/2}{}^2 \Big(1-0.0880932 \Theta_1{}^2-1.5 \Theta_1 \Theta_2+0.75 \Theta_1 \Theta_3+0.75 \Theta_1 \Theta_4-0.0880932
\Theta_2{}^2\nonumber\\
&\quad\quad\quad+0.75 \Theta_2 \Theta_3+0.75 \Theta_2 \Theta_4-0.419047 \Theta_3{}^2-1.5 \Theta_3 \Theta_4-2.40477 \Theta_4{}^2\Big)
~,~\nonumber\\
 \widetilde{m}^2_{L}& = m_{3/2}{}^2 \Big(1-2.02977 \Theta_1{}^2+0.75 \Theta_1 \Theta_2-1.5 \Theta_1 \Theta_4-0.0440466 \Theta_2{}^2-1.5 \Theta_2
\Theta_3 \nonumber\\
&\quad\quad\quad +0.286907 \Theta_3{}^2+0.75 \Theta_3 \Theta_4+0.286907 \Theta_4{}^2\Big).
\label{ssb}
\end{align}
All the above results are subject to the constraint in Eq. (\ref{Theta4}).
    \end{widetext}
	\section{Scanning Methodology and Phenomenological Constraints}
	\label{constraints}
	
We use \texttt{ISAJET 7.85}~\cite{ISAJET} to investigate the parameter space using the Metropolis-Hastings algorithm~\cite{Belanger:2009ti, Baer:2008ksh}. Among the sampled parameters are 
	\begin{itemize}
		\item $\gamma_1, \gamma_2 \in [0,1]$ (angular parameters)
		\item $\Theta_4 \in [0,1]$
		\item $m_{3/2} \in [0,15]$ TeV
		\item $\tan\beta \in [2,60]$
	\end{itemize}
	Key constraints applied:
	\begin{enumerate}
		\item \textbf{LEP bounds}~\cite{Patrignani:2016xqp}: 
		\begin{itemize}
			\item $m_{\tilde{t}_1}, m_{\tilde{b}_1} > 100$ GeV 
			\item $m_{\tilde{\tau}_1} > 105$ GeV
			\item $m_{\tilde{\chi}^\pm_1} > 103$ GeV
		\end{itemize}
		\item \textbf{Higgs mass}~\cite{Khachatryan:2016vau}: $122 \leq m_h \leq 128$ GeV (allowing for theoretical uncertainties~\cite{Slavich:2020zjv, Allanach:2004rh})
		\item \textbf{B-physics}~\cite{CMS:2014xfa, Amhis:2014hma}:
		\begin{itemize}
			\item $1.6 \times 10^{-9} \leq \text{BR}(B_s \to \mu^+\mu^-) \leq 4.2 \times 10^{-9}$
			\item $2.99 \times 10^{-4} \leq \text{BR}(b \to s\gamma) \leq 3.87 \times 10^{-4}$
			\item $0.70 \times 10^{-4} \leq \text{BR}(B_u \to \tau\nu) \leq 1.5 \times 10^{-4}$
		\end{itemize}
		\item \textbf{LHC sparticle limits}~\cite{ATLAS:2017mjy, Vami:2019slp, CMS:2017okm}: $m_{\tilde{g}} \gtrsim 2.2$ TeV and $m_{\tilde{q}} \gtrsim 2$ TeV.
		\item \textbf{Dark matter relic density}~\cite{Planck:2018nkj}: $0.114 \leq \Omega_{\rm CDM}h^2 \leq 0.126$
	\end{enumerate}
    
Only solutions with neutralino LSP that satisfy radiative electroweak symmetry breaking (REWSB) are kept. Compatibility with DM observations and collider phenomenology is given priority in the parameter space exploration.

    
	\section{Numerical Results and Discussions}
	\label{results}
In Fig.~\ref{input_params1}, we display the $\Theta_1-\Theta_2$, $\Theta_1-\Theta_3$, and $\Theta_3-\Theta_2$ planes. The left and right panels correspond to scenarios with $\mu < 0$ and $\mu > 0$, respectively. The color scheme is defined as follows:

\begin{itemize}
    \item \textbf{Grey points}: Represent solutions that satisfy the conditions for radiative electroweak symmetry breaking (REWSB) and feature a neutralino as the lightest supersymmetric particle (LSP).
    
    \item \textbf{Blue points}: A subset of grey points that, in addition to satisfying REWSB, are consistent with current experimental constraints, including the LHC bounds on sparticle masses, B-physics observables, and the Higgs boson mass measurement. These points predict an overabundant relic density.
    
    \item \textbf{Green points}: A subset of blue points corresponding to solutions with an underabundant relic density.
    
    \item \textbf{Red points}: A subset of green points that yield a relic density within the $5\sigma$ range of the Planck 2018 measurements, satisfying the observed DM abundance.
\end{itemize}

	\begin{widetext}
	
\begin{figure}[h!]
	\centering \includegraphics[width=7.90cm]{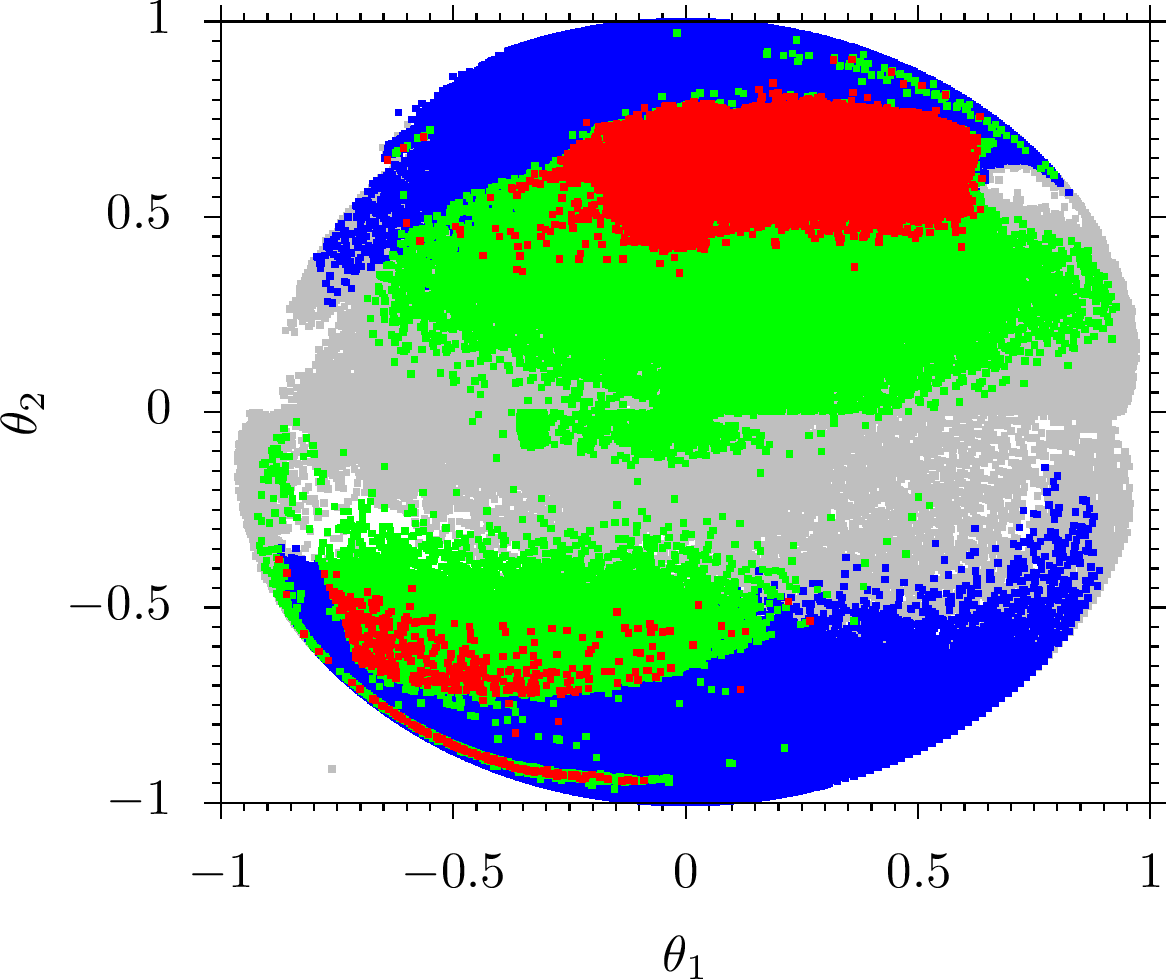}
    \centering \includegraphics[width=7.90cm]{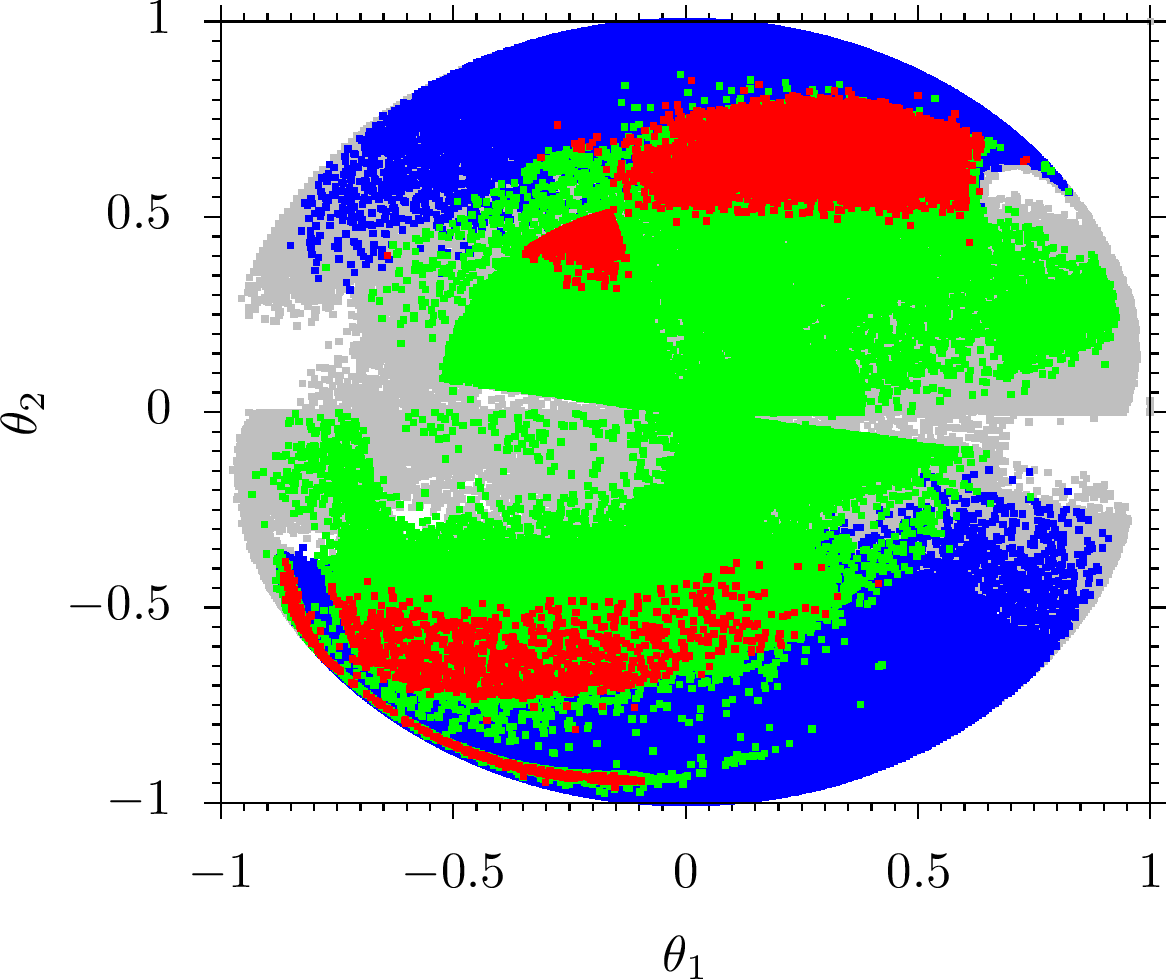}
	\centering \includegraphics[width=7.90cm]{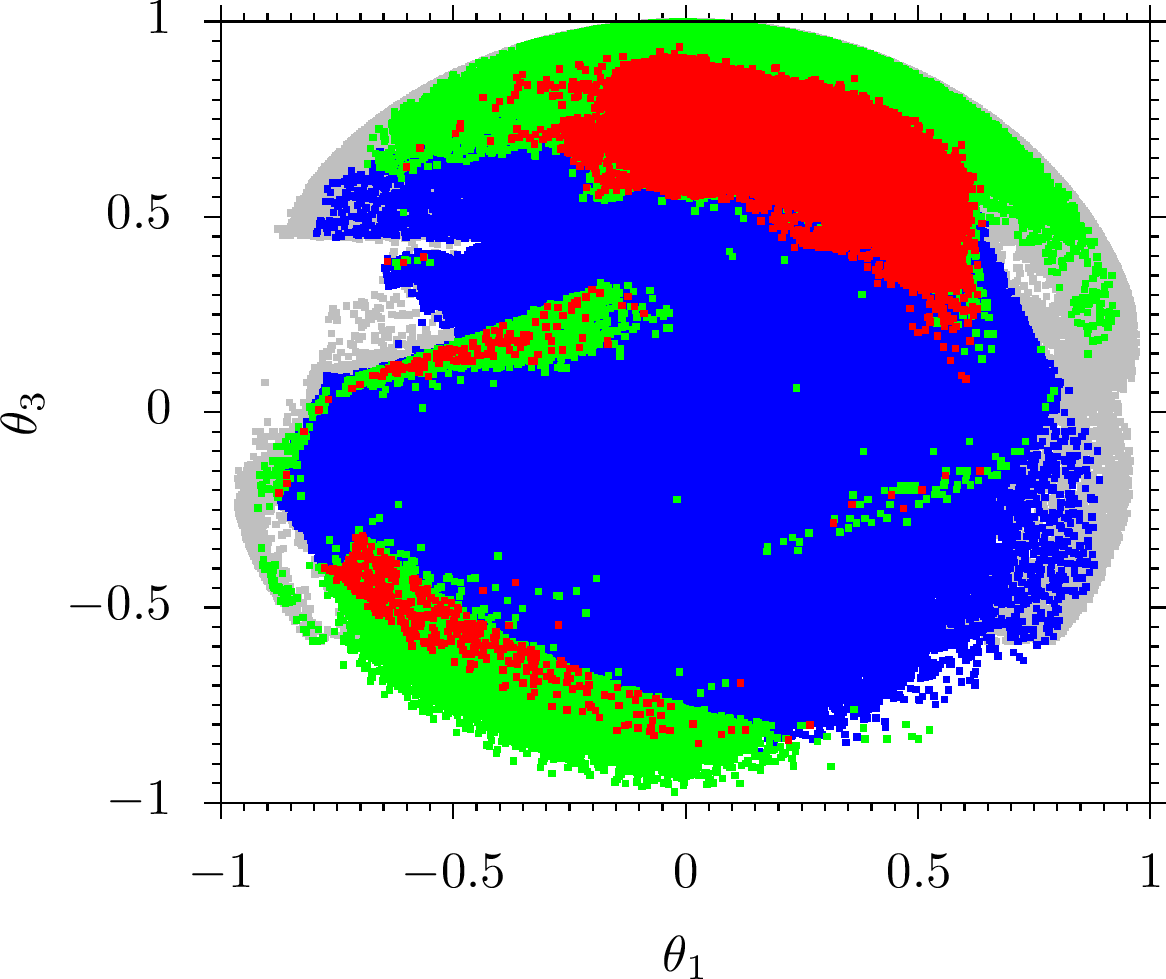}
    \centering \includegraphics[width=7.90cm]{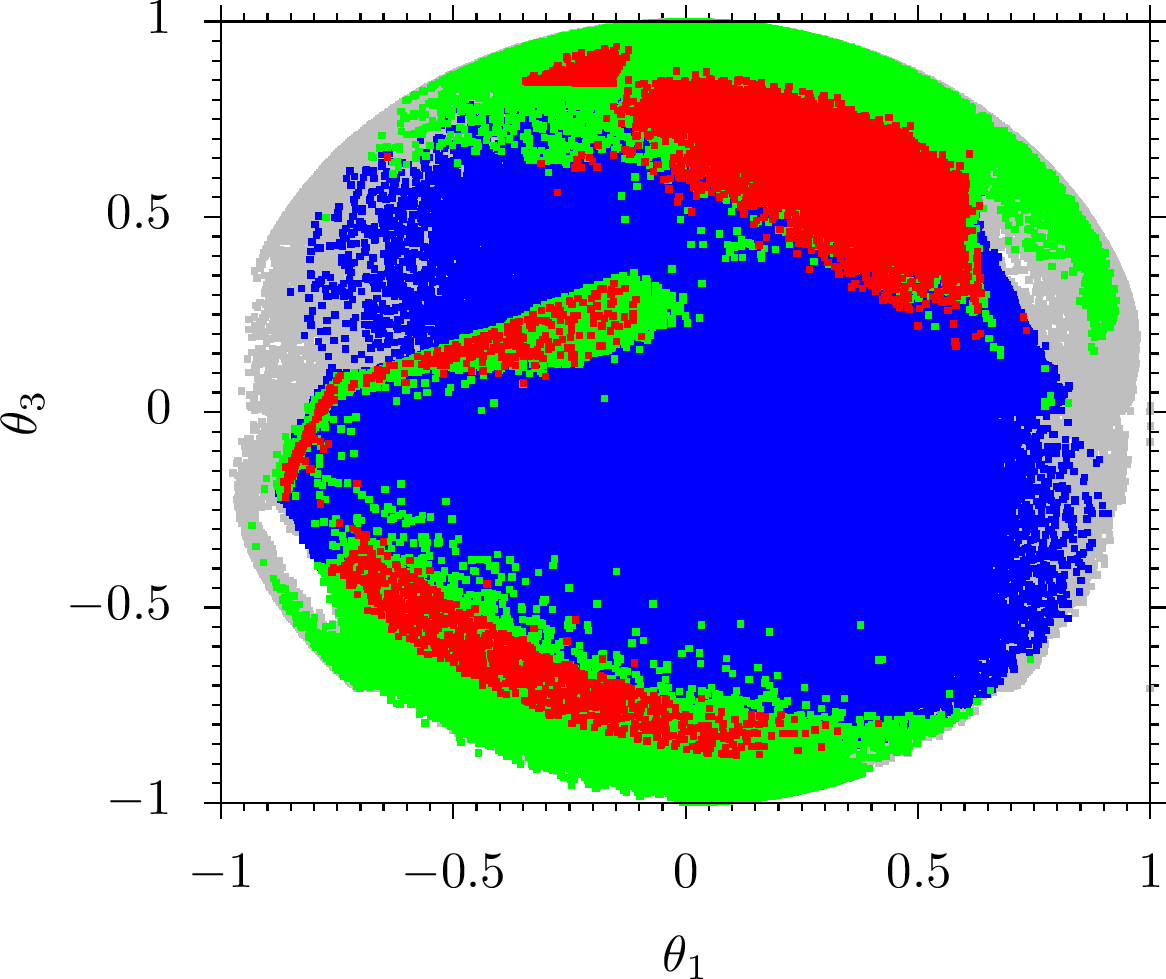}
	\centering \includegraphics[width=7.90cm]{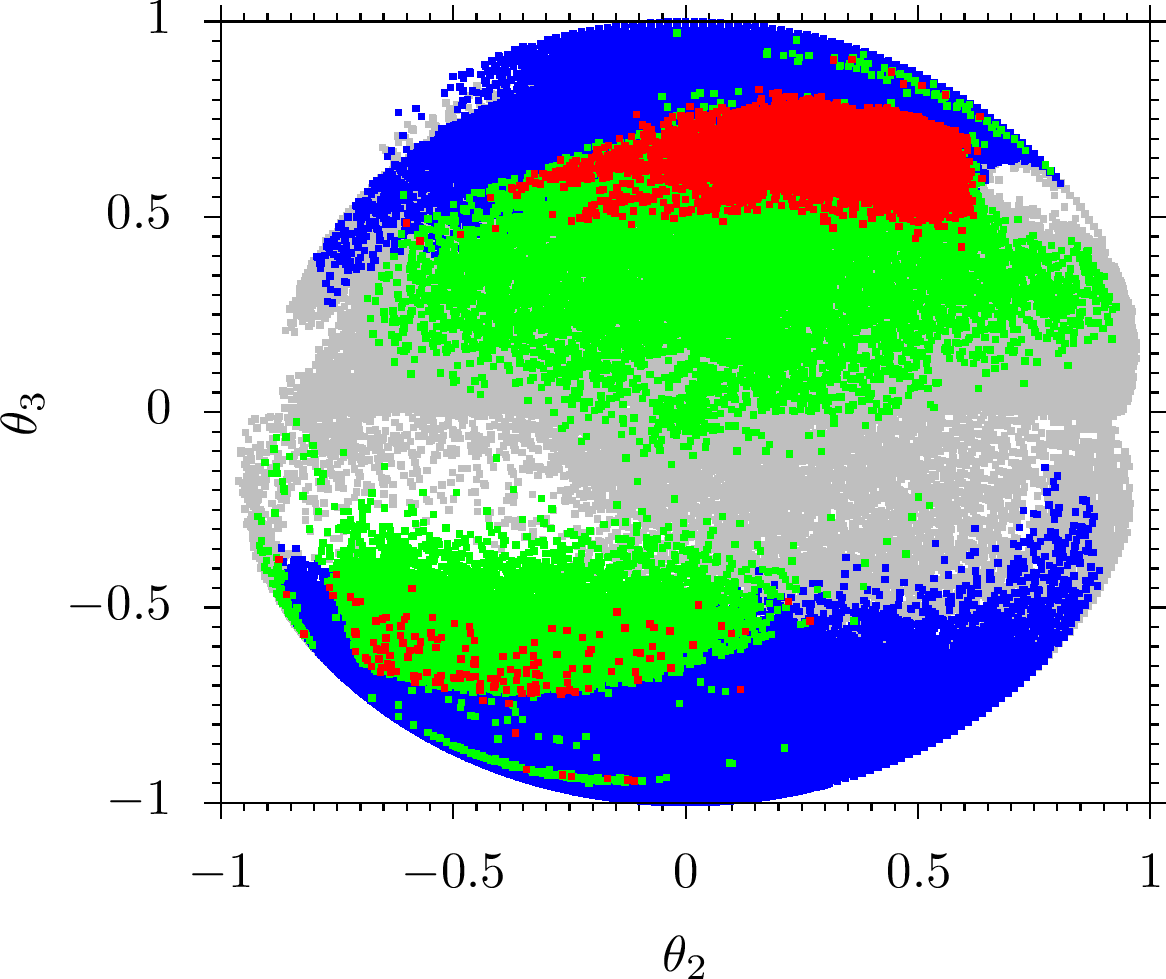}
    \centering \includegraphics[width=7.90cm]{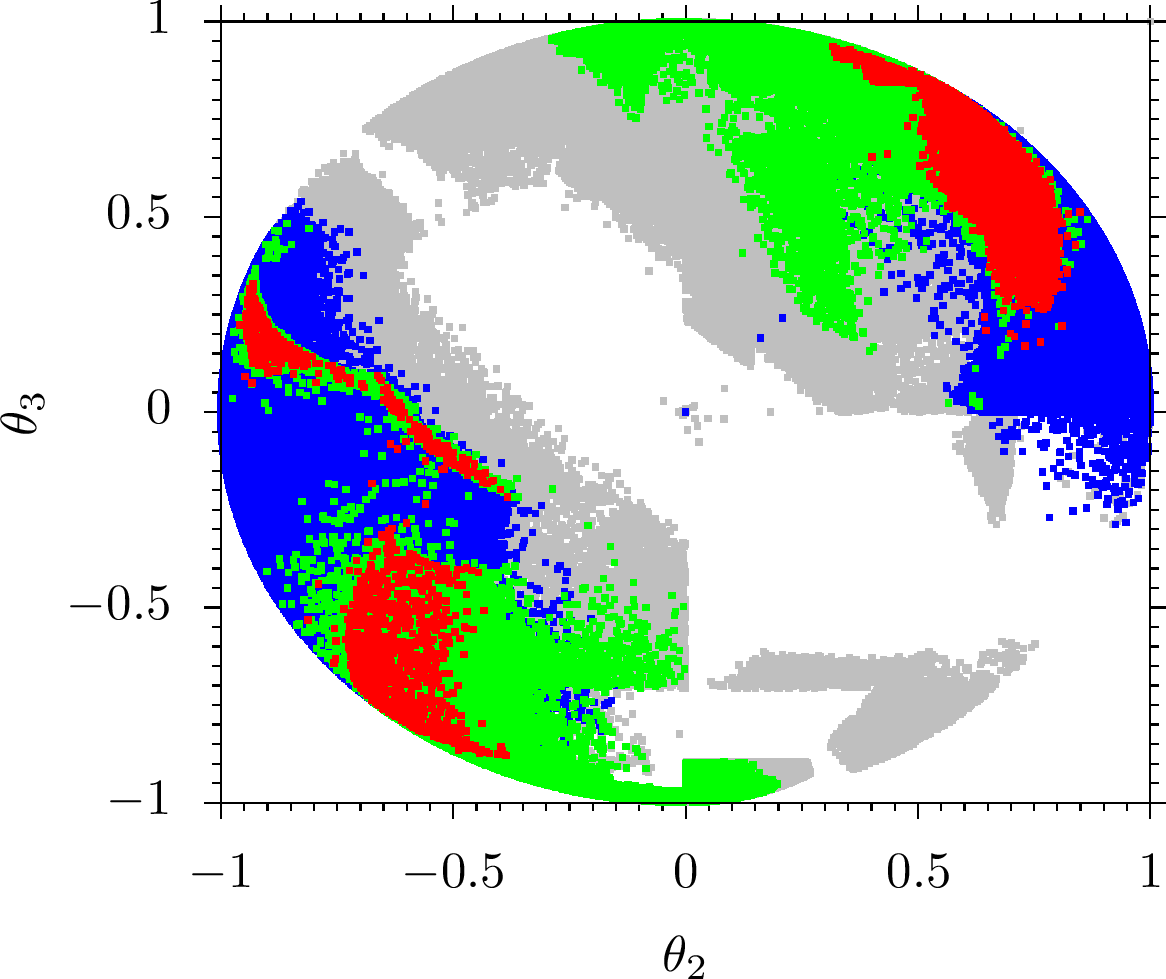}
    \caption{
			Plots in $\Theta_1-\Theta_2$, $\Theta_1-\Theta_3$ and $\Theta_3-\Theta_2$ 
			planes for $\mu <0$ (left panels) and $\mu > 0$ (right panels). Grey points complete REWSB and offer LSP neutralino. Blue points are a subset of gray points that meet all LHC SUSY particle mass bounds, B-physics, as well as LHC Higgs mass bounds, indicating over-saturated relic density.  Green points are a subset of blue points that indicate under-saturated relic density. Finally, red points are a subset of green points and reflect the saturated relic density bound (Planck 2018 5$\sigma$ bounds).
		}
		\label{input_params1}
\end{figure}

\begin{figure}[h!]
	\centering \includegraphics[width=7.90cm]{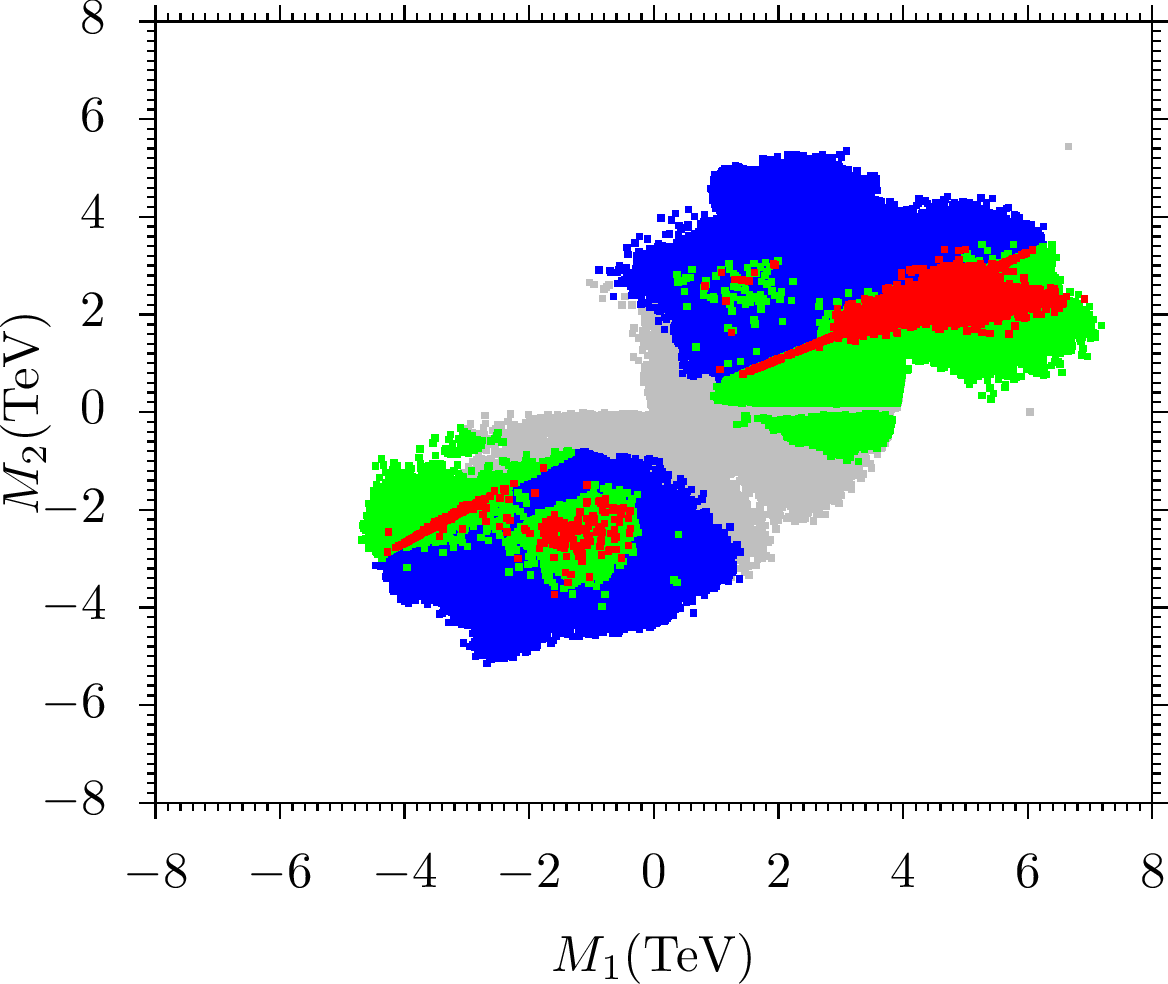}
    \centering \includegraphics[width=7.90cm]{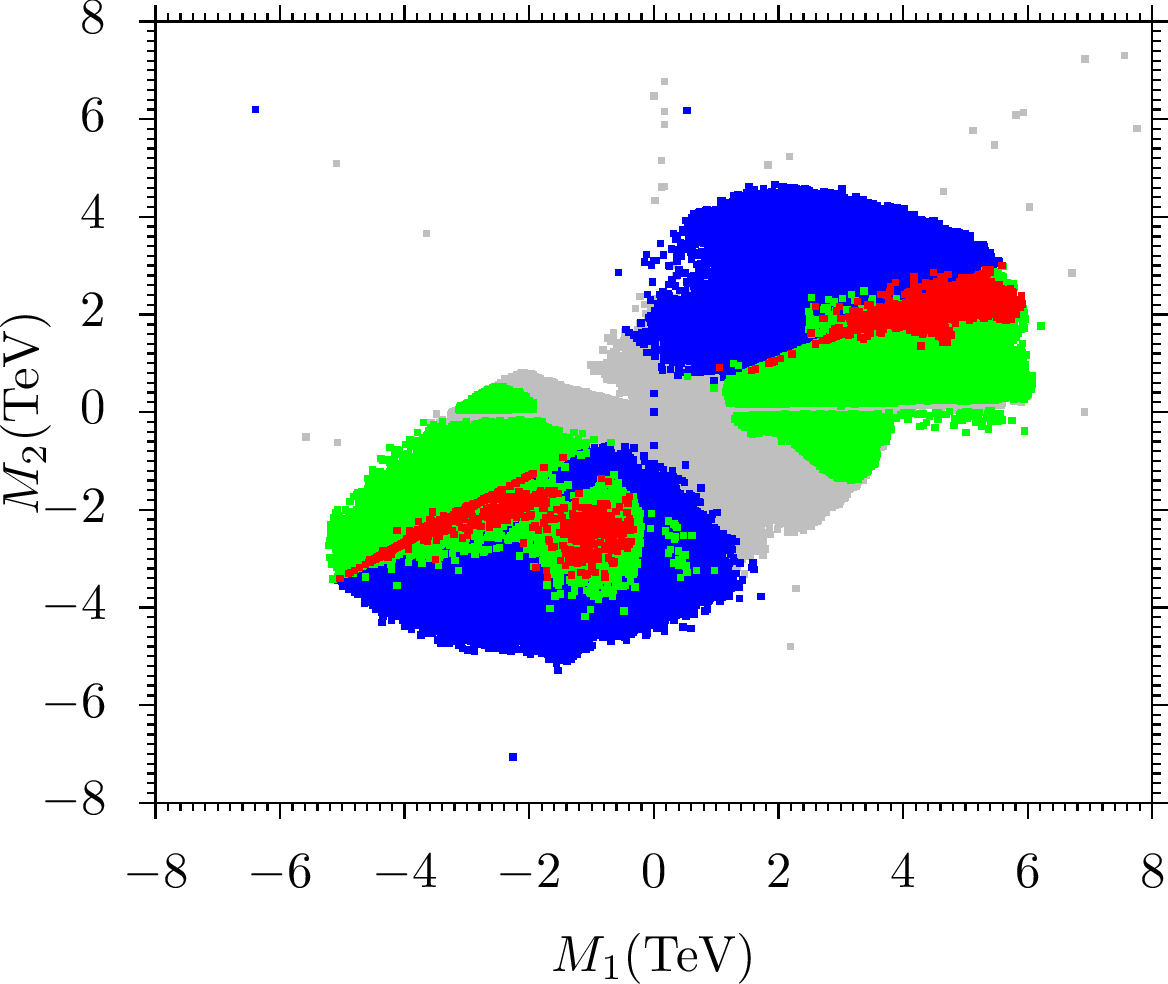}
	\centering \includegraphics[width=7.90cm]{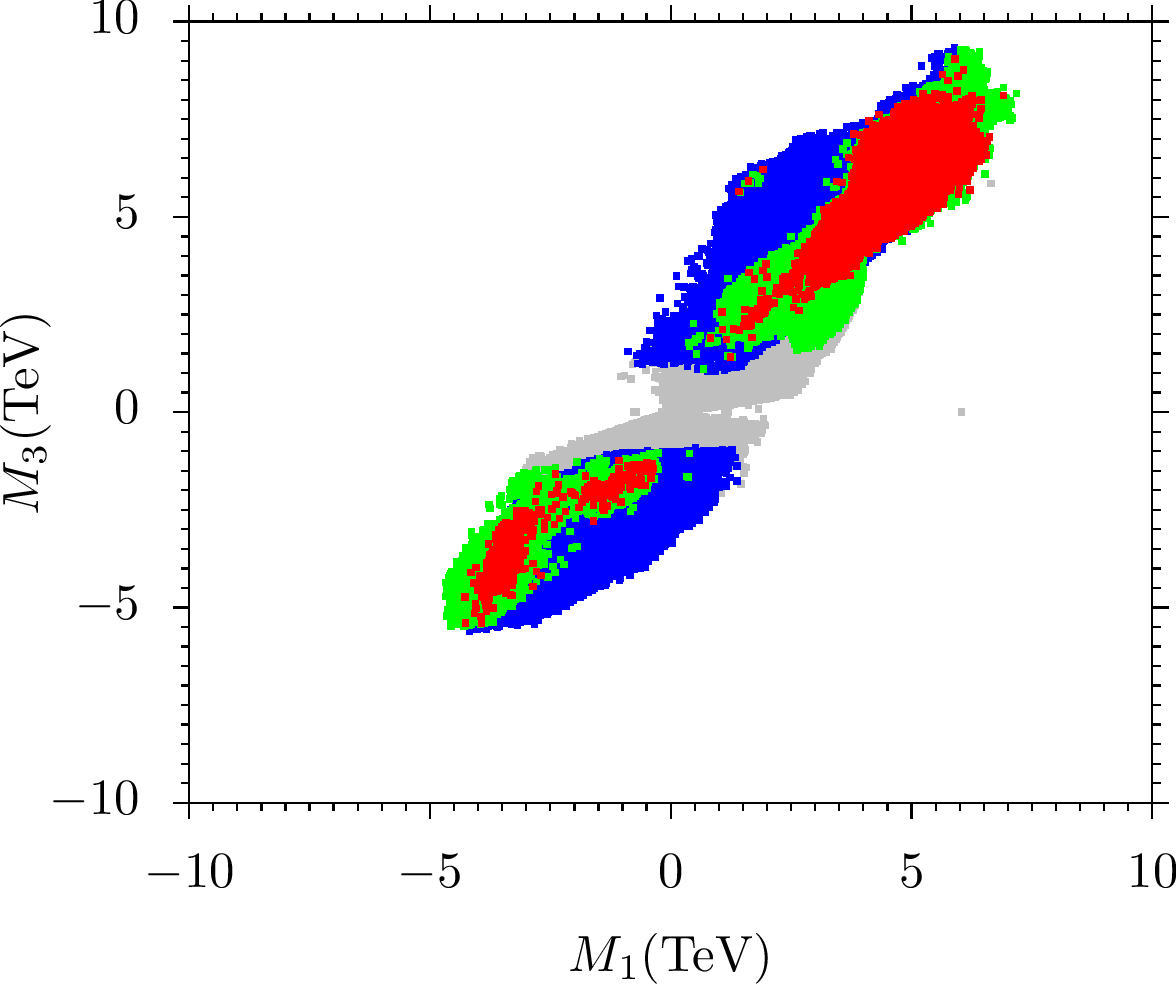}
    \centering \includegraphics[width=7.90cm]{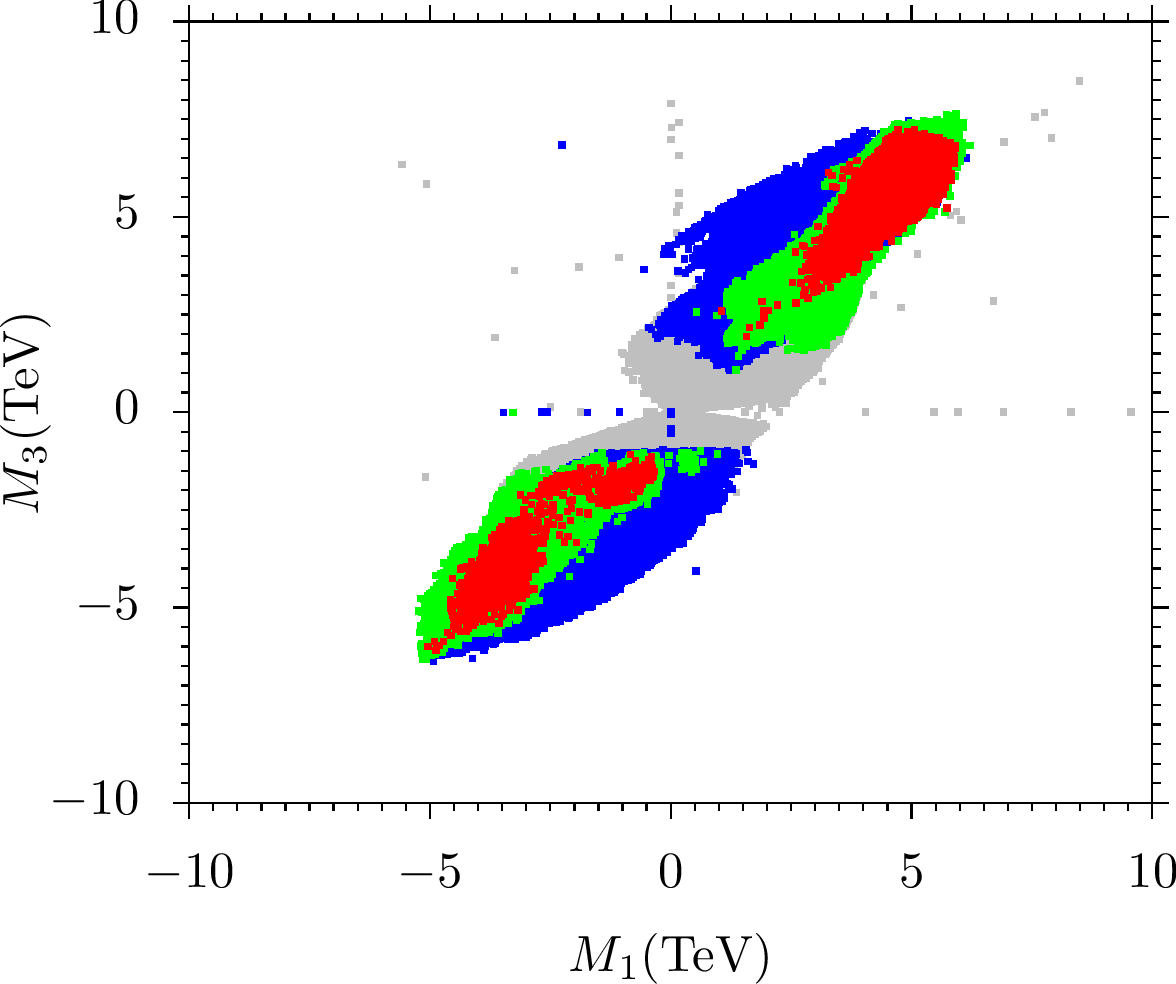}
	\centering \includegraphics[width=7.90cm]{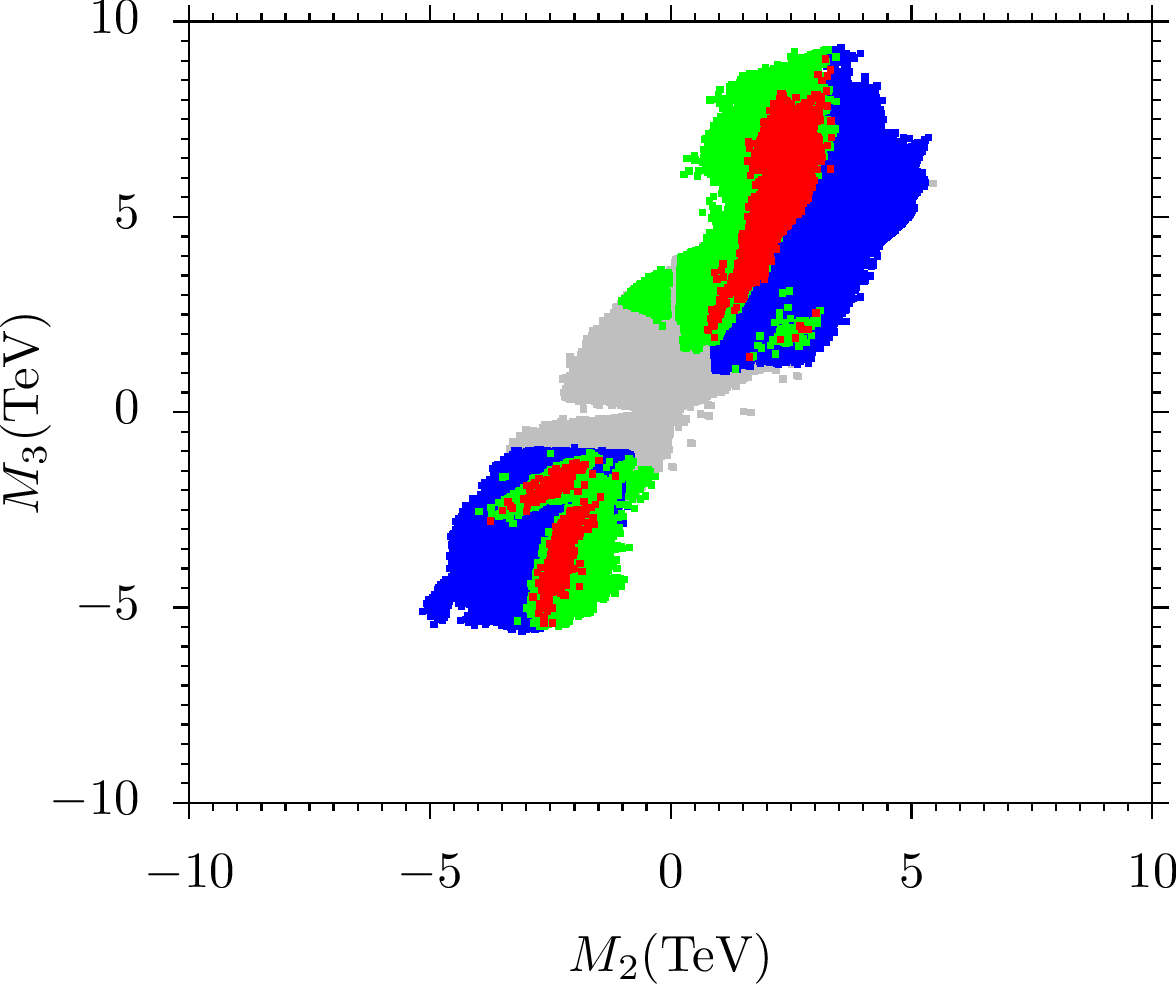}
    \centering \includegraphics[width=7.90cm]{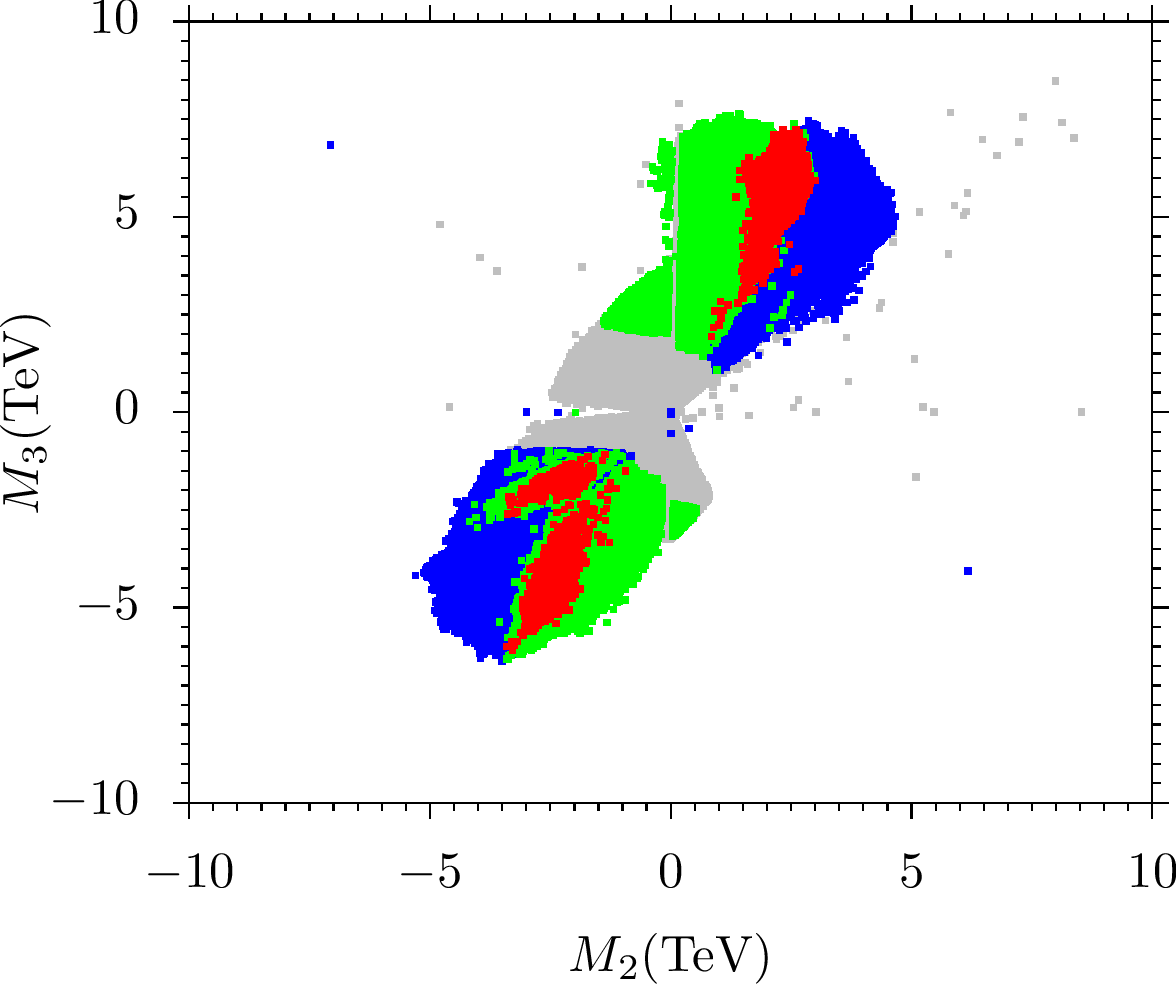}
    \caption{Plots in $M_1-M_2$, $M_1-M_3$ and $M_3-M_2$ planes for $\mu <0$ (left panels) and $\mu > 0$ (right panels). Grey points complete REWSB and offer LSP neutralino. Blue points are a subset of gray points that meet all LHC SUSY particle mass bounds, B-physics, as well as LHC Higgs mass bounds, indicating an over-saturated relic density.  Green points are a subset of  blue points that indicate under-saturated relic density. Finally, red points are a subset of green points and reflect the saturated relic density bound (Planck 2018 5$\sigma$ bounds).
		}
		\label{funda_params2}
\end{figure}

	\begin{figure}[t]
	\centering \includegraphics[width=7.90cm]{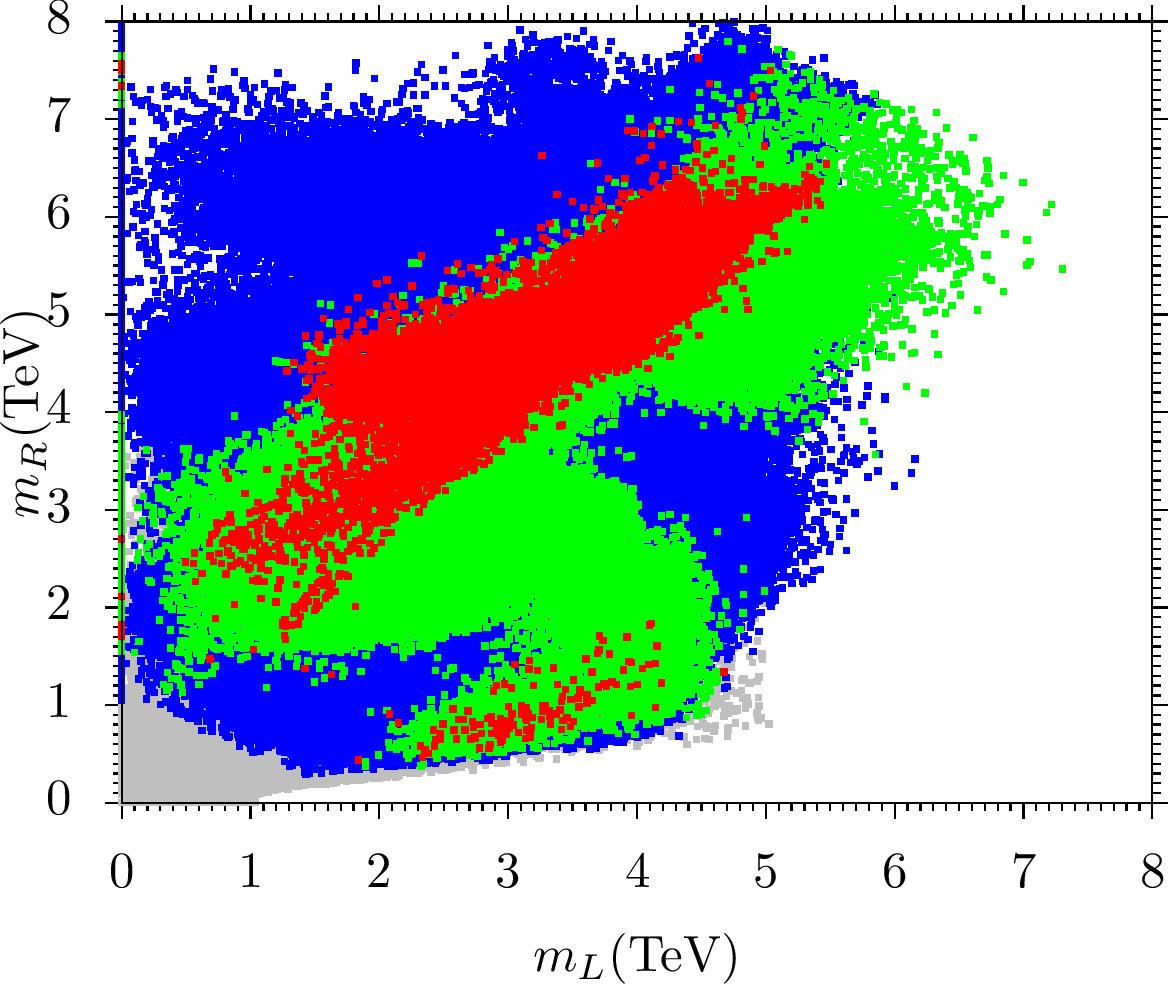}
	\centering \includegraphics[width=7.90cm]{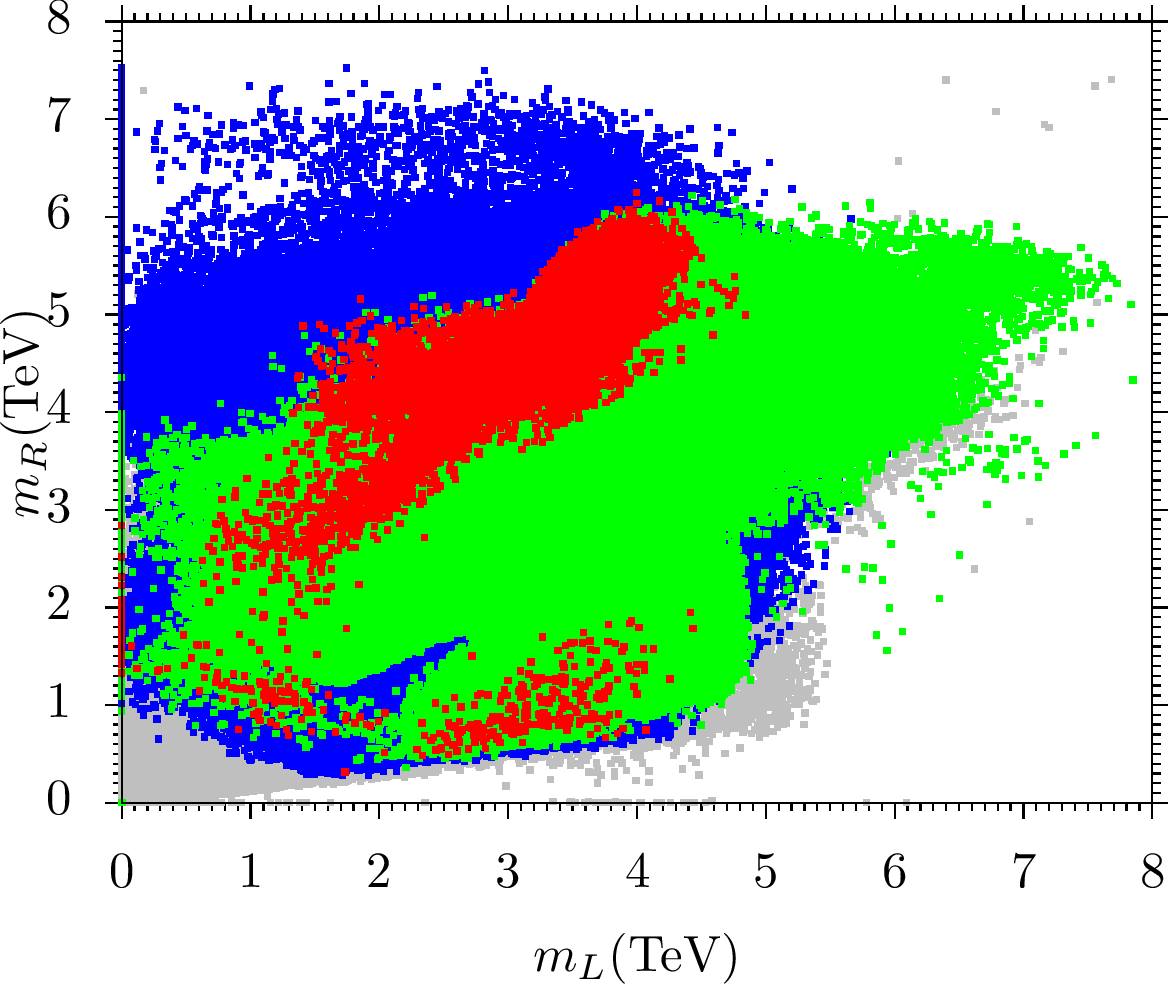}
	\centering \includegraphics[width=7.90cm]{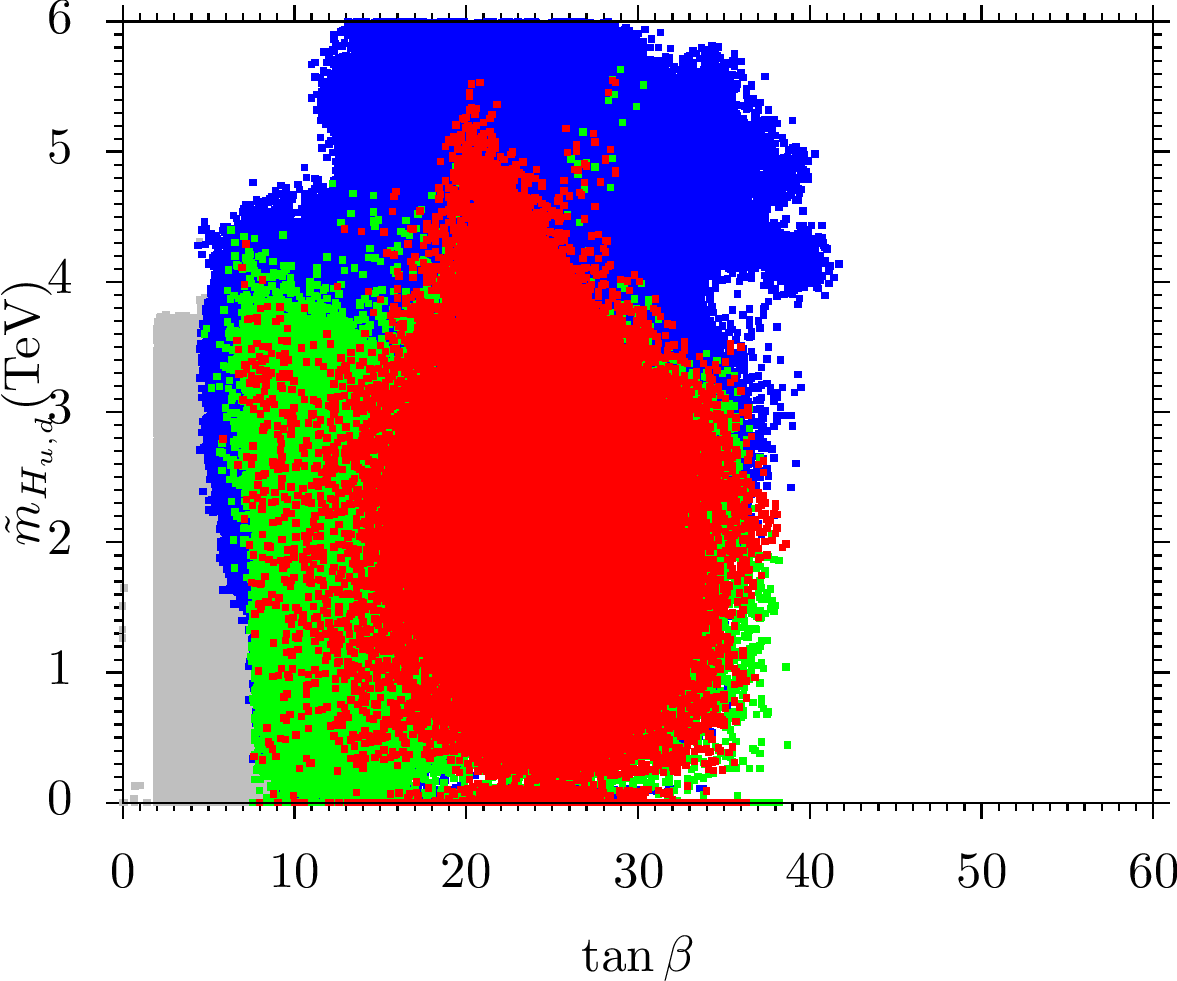}
    \centering \includegraphics[width=7.90cm]{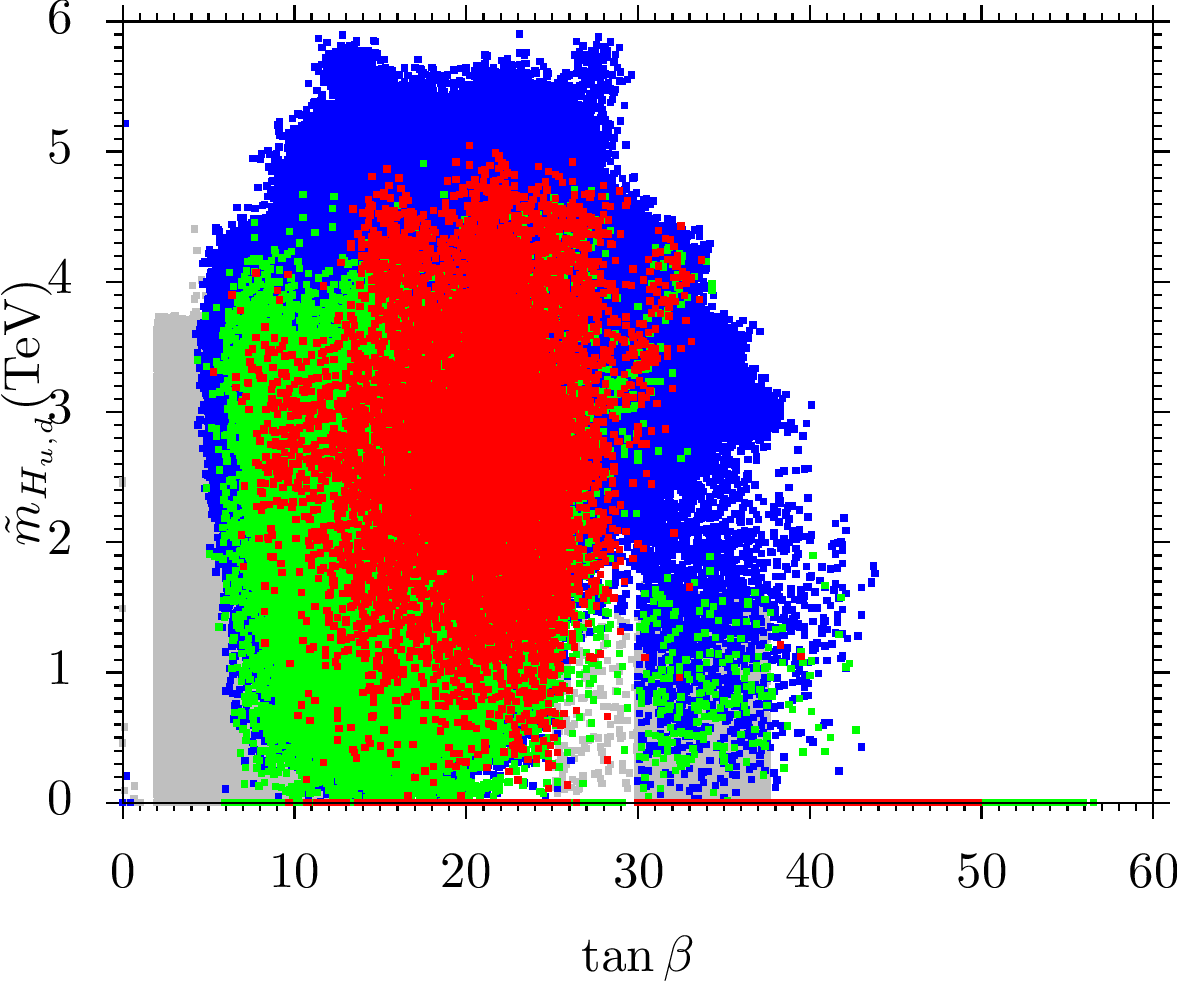}
    \centering \includegraphics[width=7.90cm]{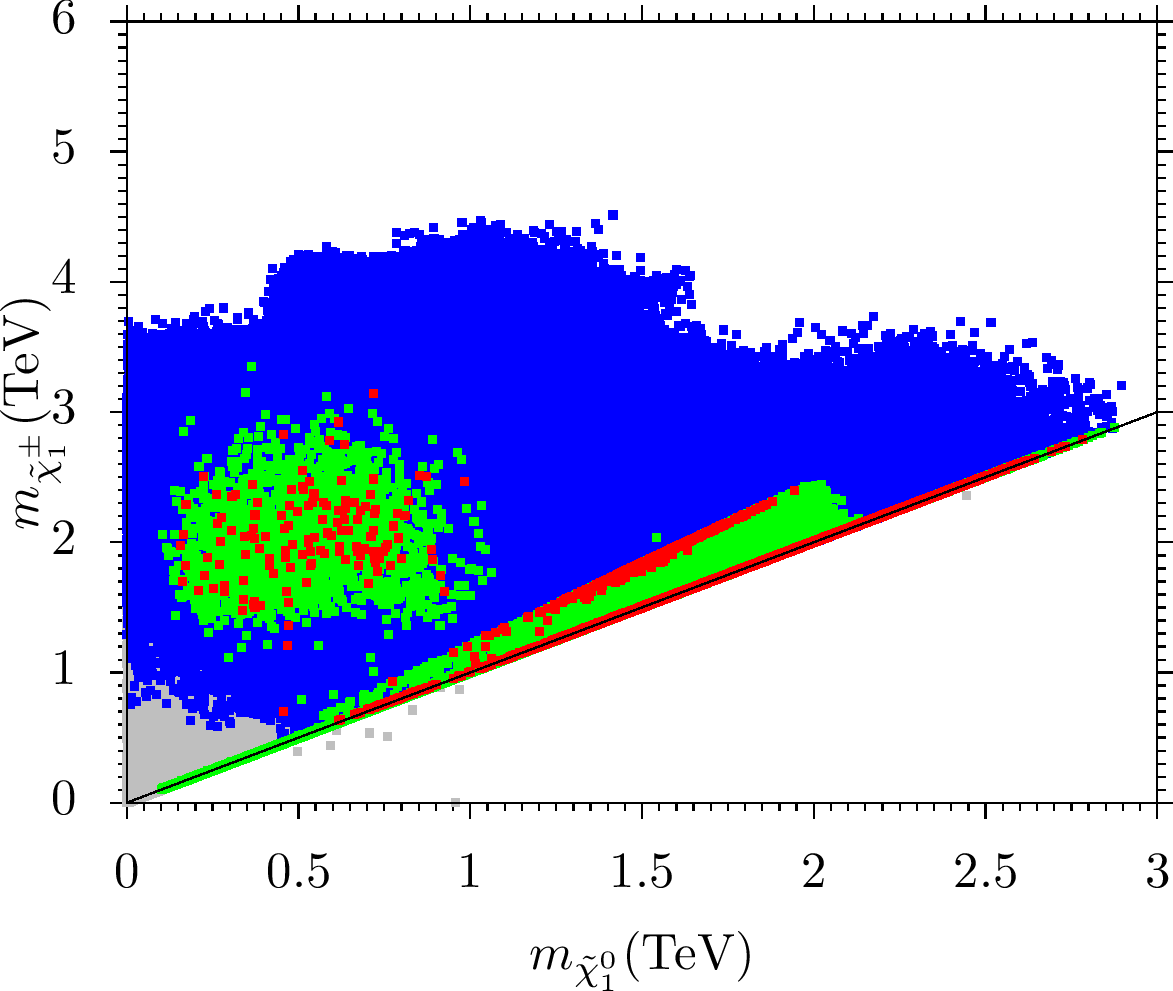}
    \centering \includegraphics[width=7.90cm]{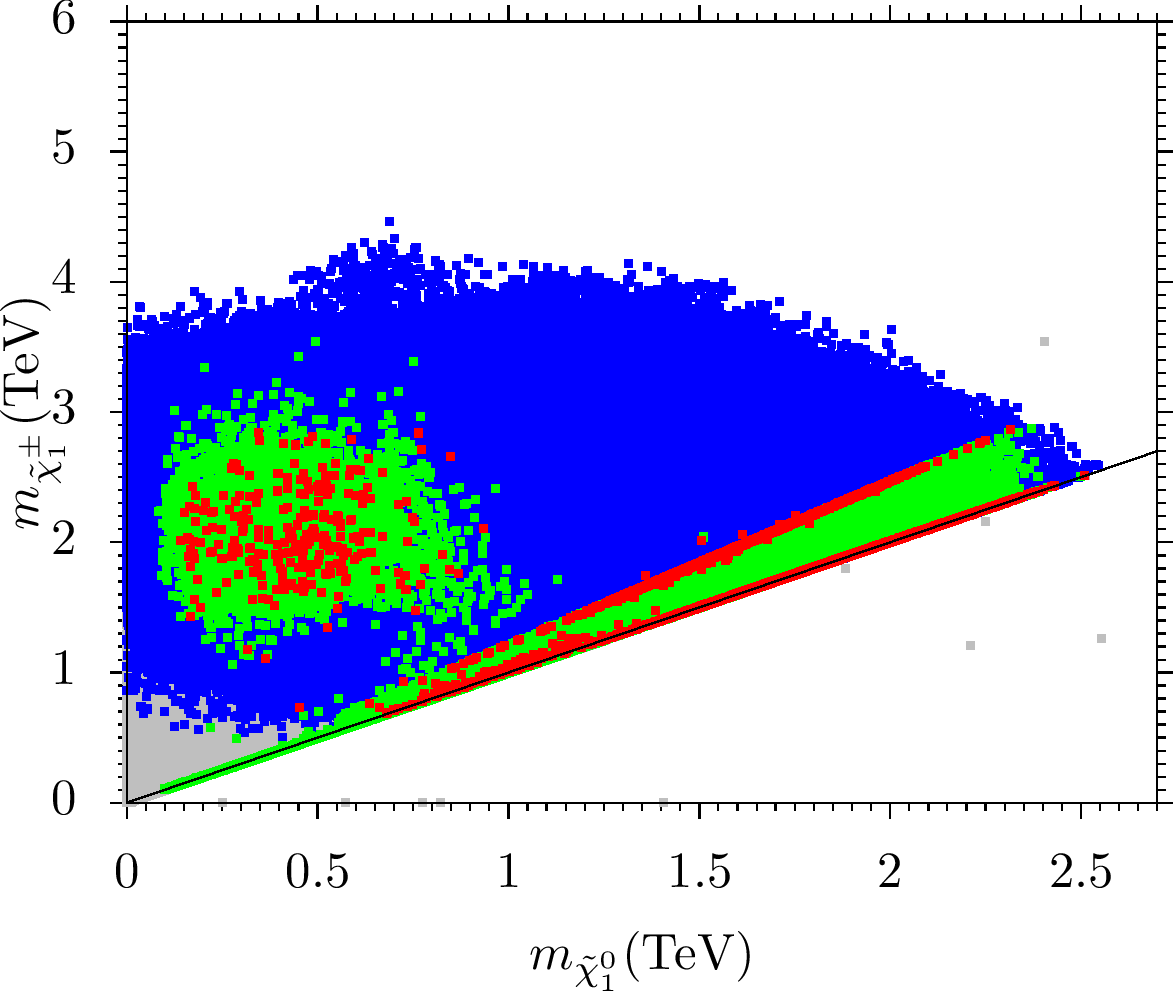}
    \caption{Plots in $m_L-m_R$, $\tan\beta-m_{H_{u,d}}$ and $m_{\tilde{\chi}_1^0}$ - $m_{\tilde{\chi}_1^{\pm}}$ planes for $\mu <0$ (left panels) and $\mu > 0$ (right panels). Grey points complete REWSB and offer LSP neutralino. Blue points are a subset of gray points that meet all LHC SUSY particle mass bounds, B-physics, as well as LHC Higgs mass bounds, indicating over-saturated relic density.  Green points are a subset of  blue points that indicate under-saturated relic density. Finally, red points are a subset of green points and reflect the saturated relic density bound (Planck 2018, 5$\sigma$ bounds).
		}
		\label{funda_params3}
\end{figure}

	\begin{figure}[t]
	     \centering \includegraphics[width=7.90cm]{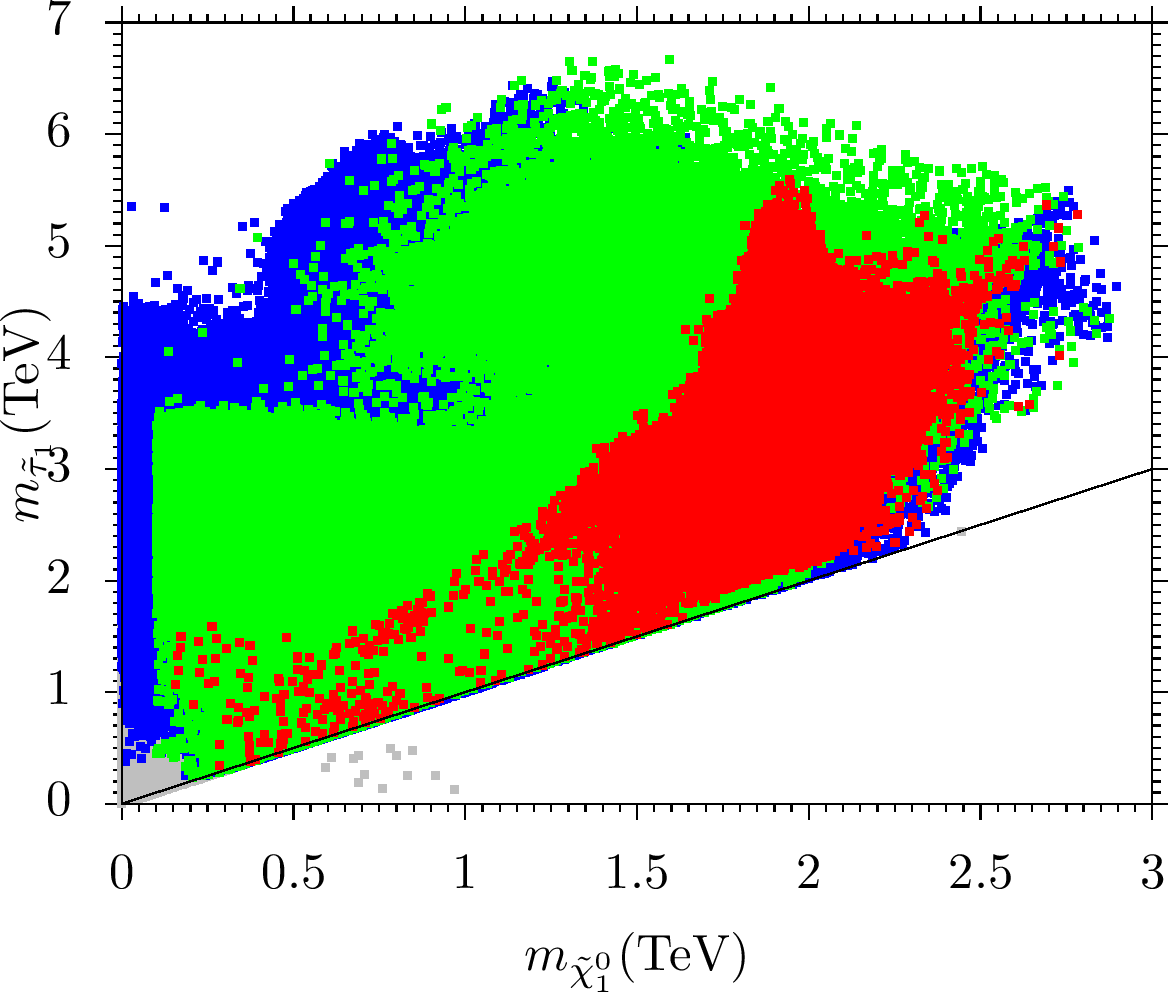}
      \centering \includegraphics[width=7.90cm]{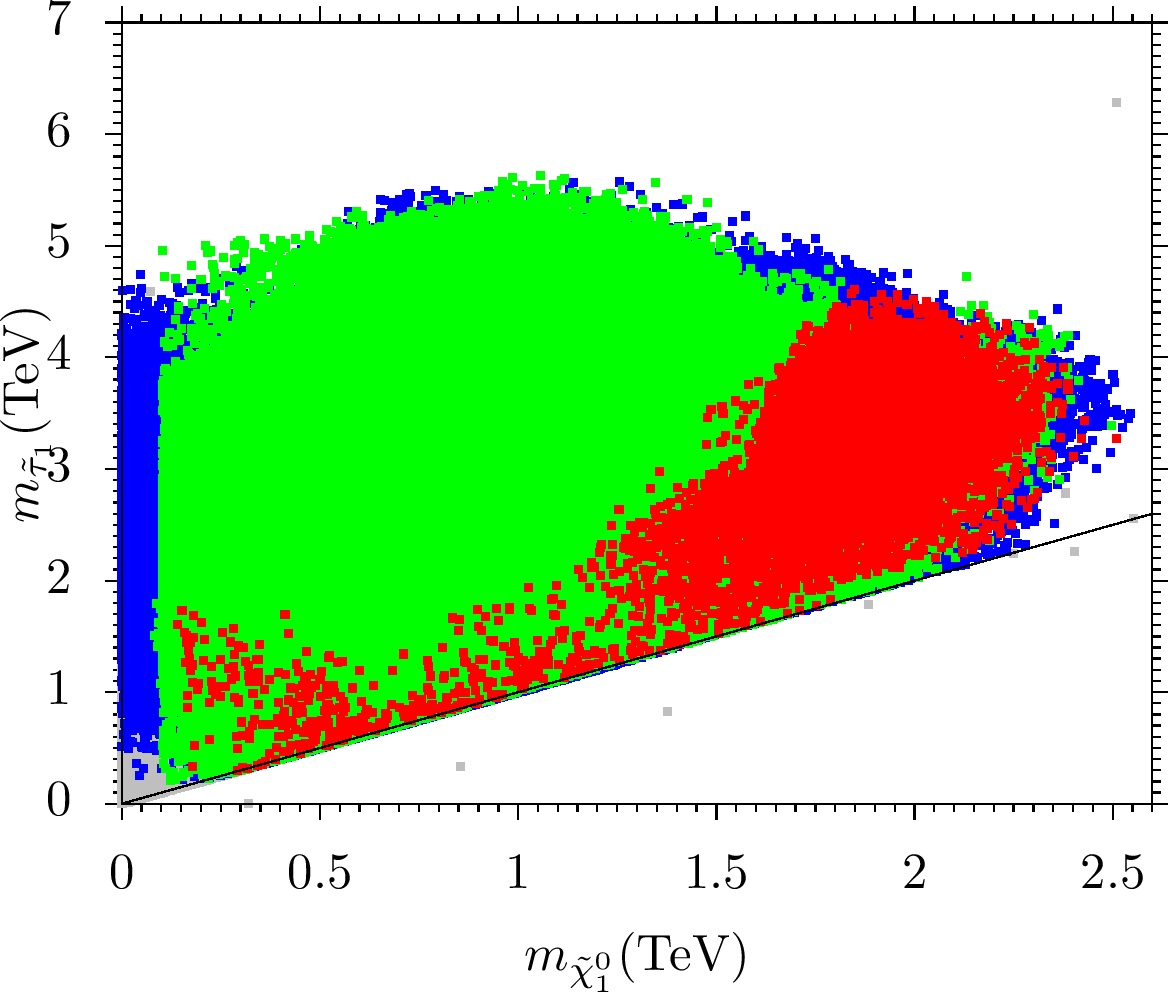}
      \centering \includegraphics[width=7.90cm]{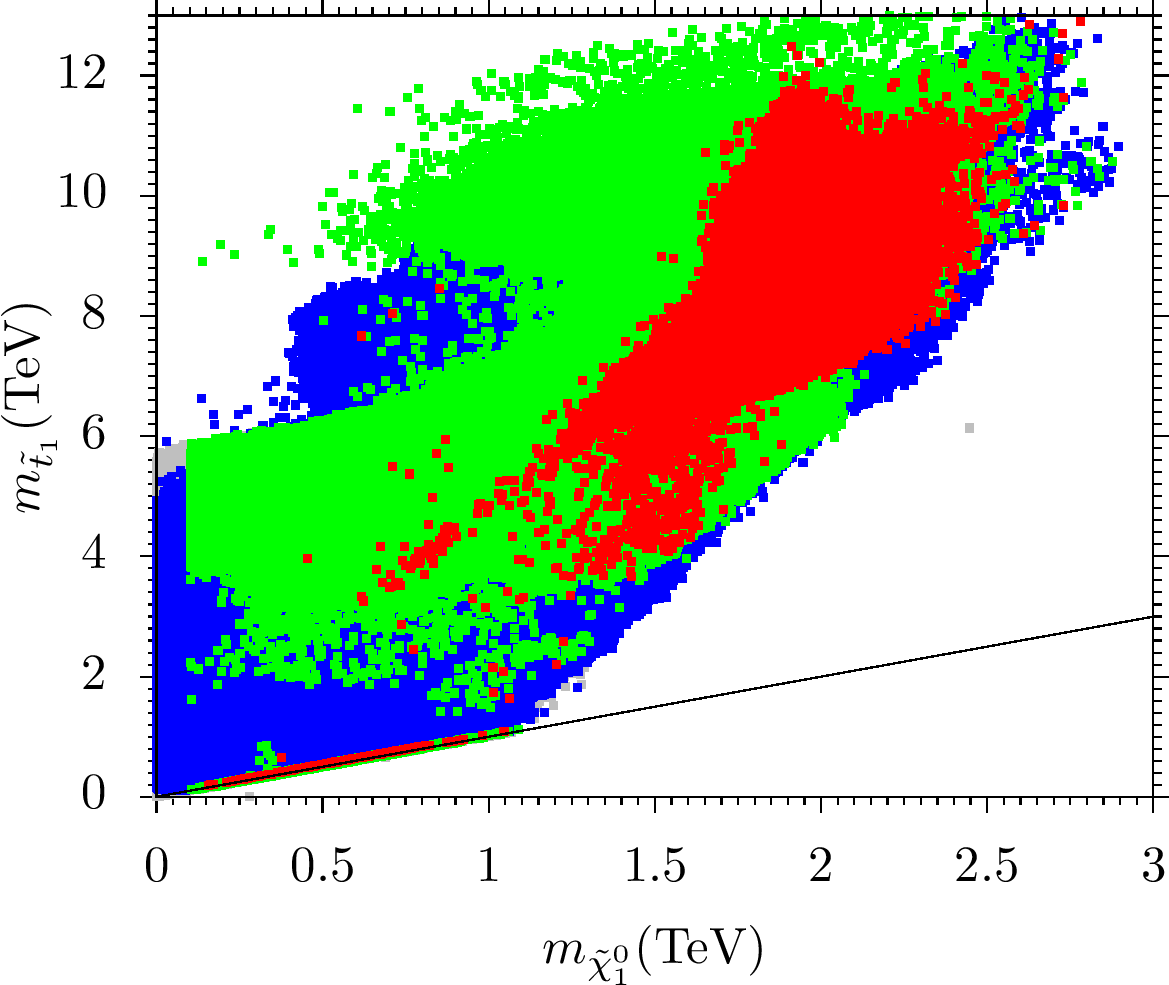}
	\centering \includegraphics[width=7.90cm]{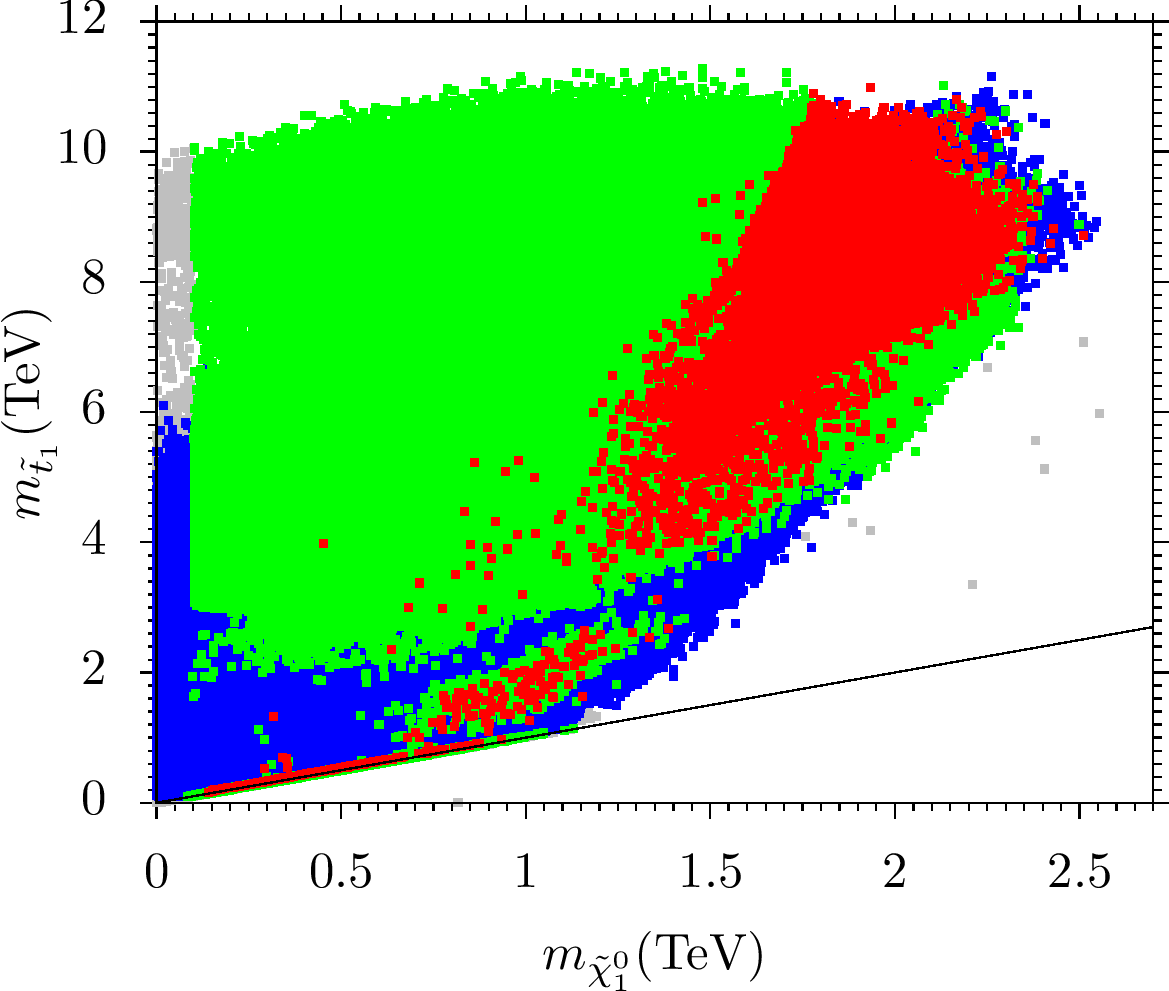}
     \centering \includegraphics[width=7.90cm]{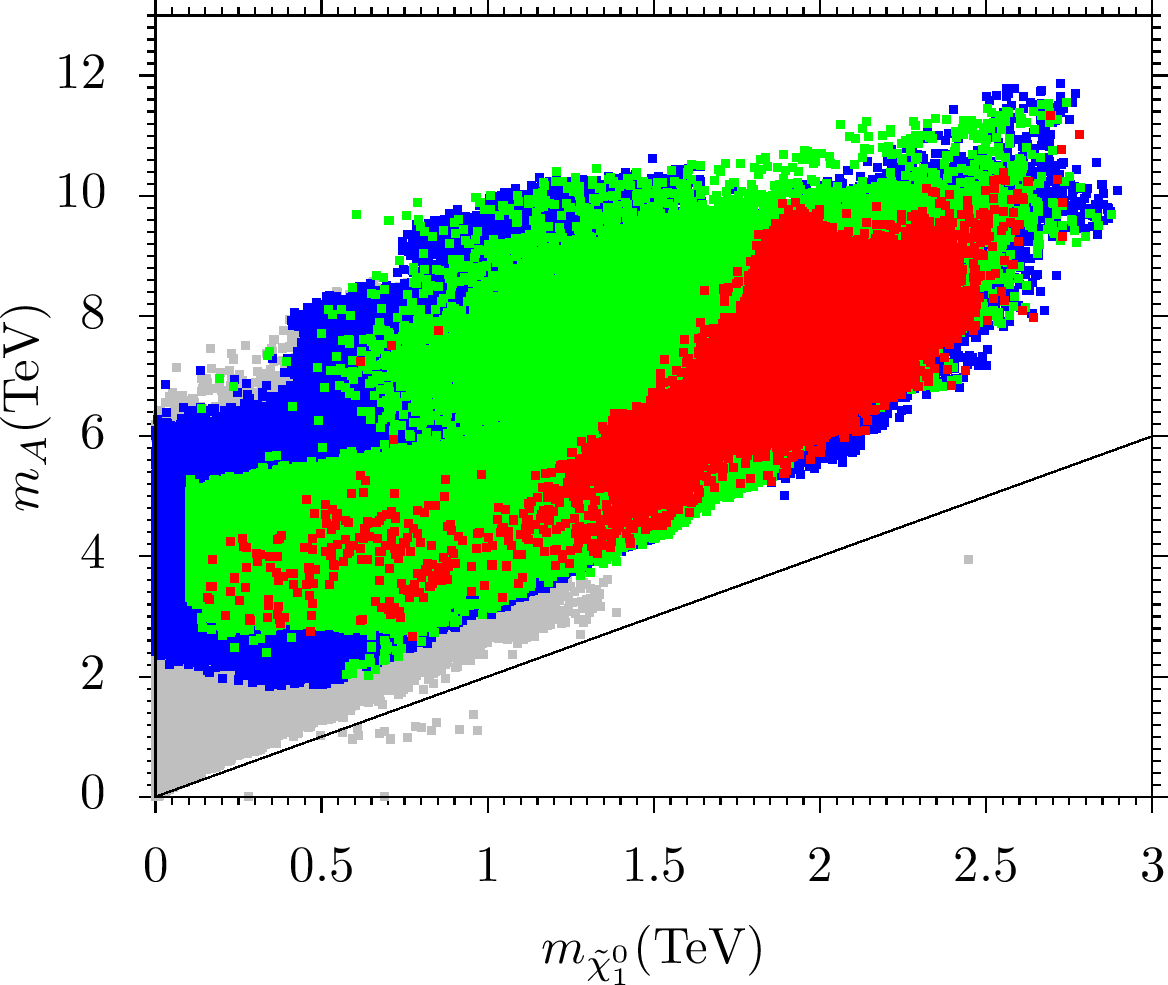}
      \centering \includegraphics[width=7.90cm]{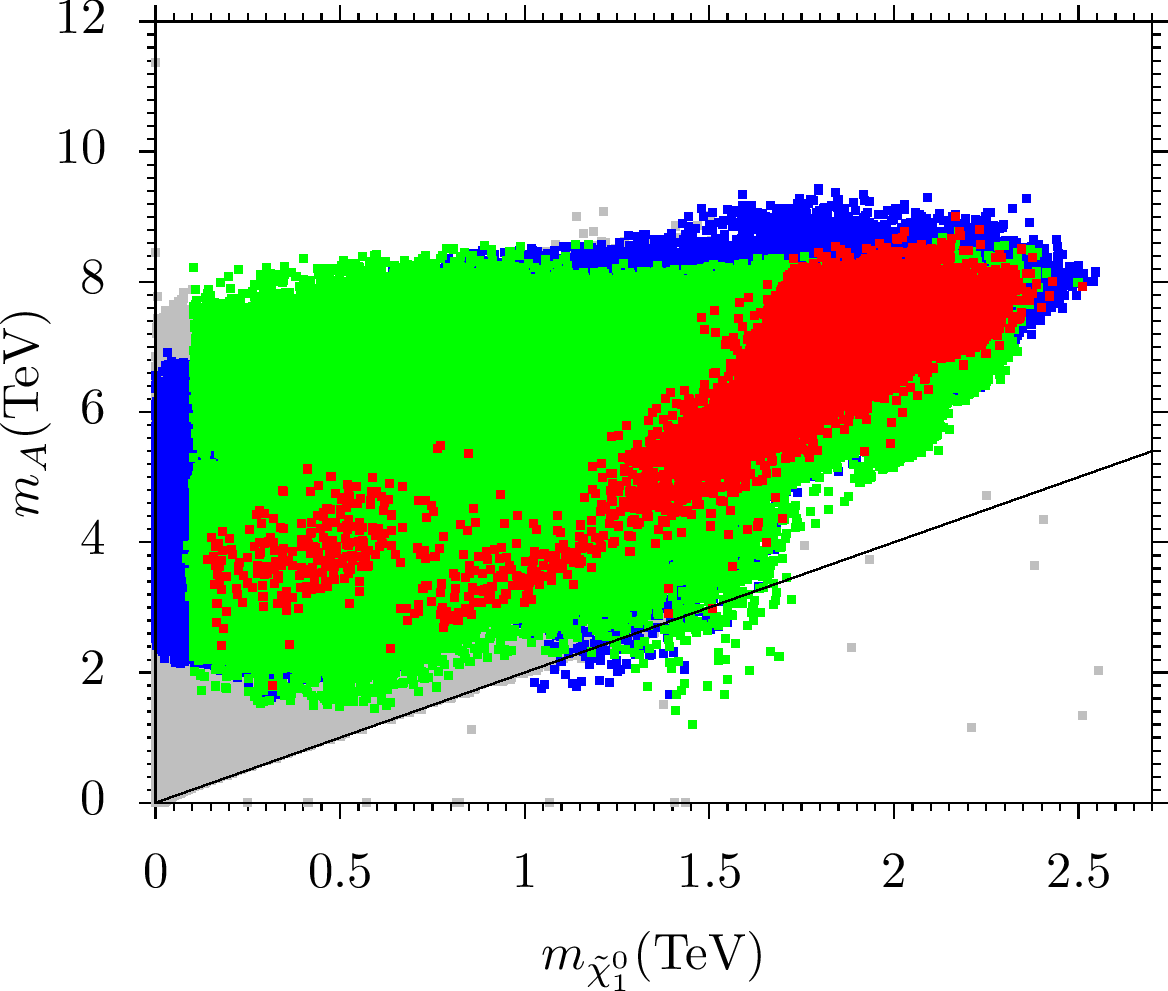}
   \caption{Plots in $m_{\tilde{\chi}_1^0}$ - $m_{\tilde{\tau}_1}$, $m_{\tilde{\chi}_1^0}$ - $m_{\tilde{t}_1}$, and $m_{\tilde{\chi}_1^0}$ - $m_A$ planes for $\mu <0$ (left panels) and $\mu > 0$ (right panels). Grey points complete REWSB and offer LSP neutralino. Blue points are a subset of gray points that meet all LHC SUSY particle mass bounds, B-physics, as well as LHC Higgs mass bounds, indicating an over-saturated relic density.  Green points are a subset of  blue points that indicate under-saturated relic density. Finally, red points are a subset of green points and reflect the saturated relic density bound (Planck 2018, 5$\sigma$ bounds).
		}
		\label{delew}
\end{figure}

    \begin{figure}[h!]
	\centering \includegraphics[width=7.90cm]{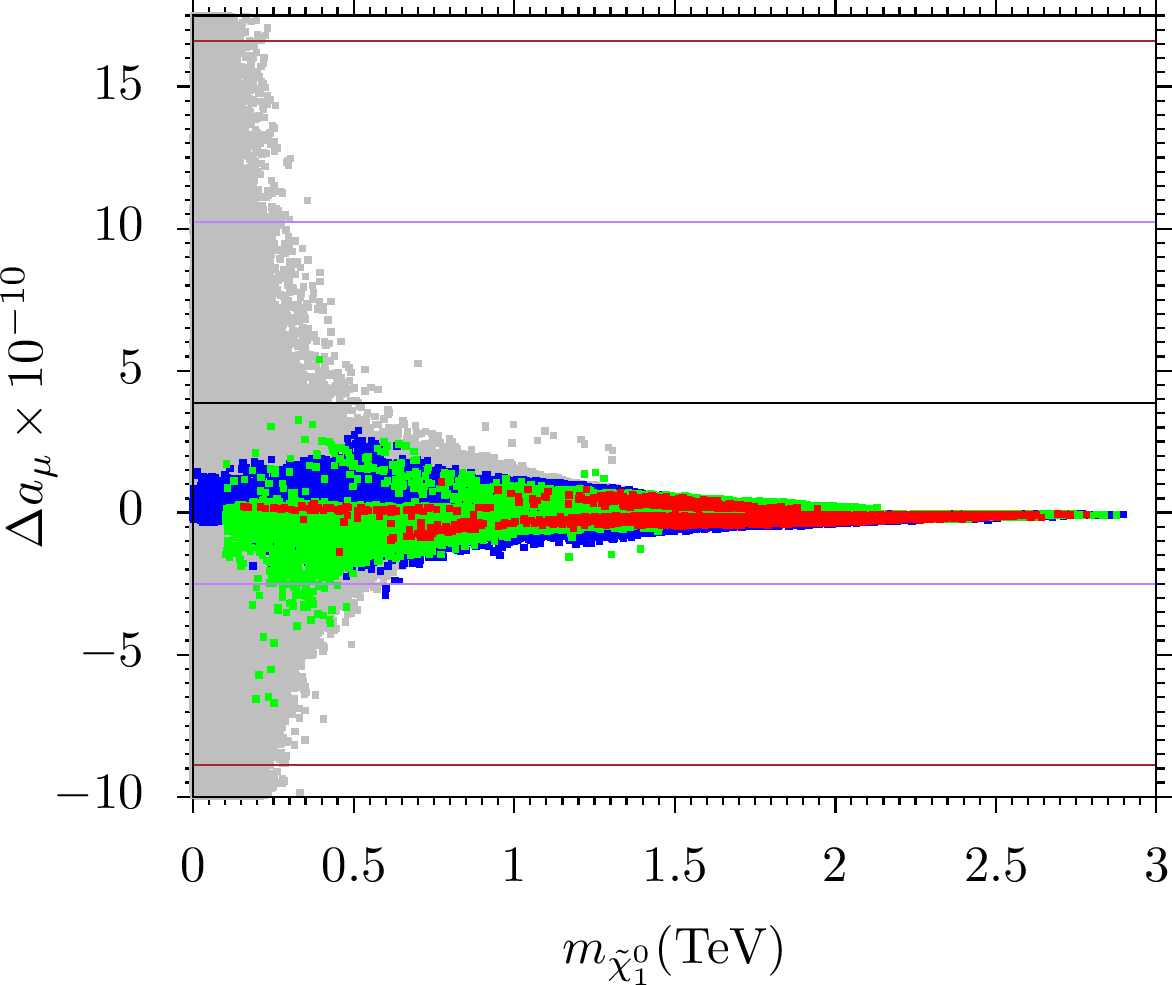}
    \centering \includegraphics[width=7.90cm]{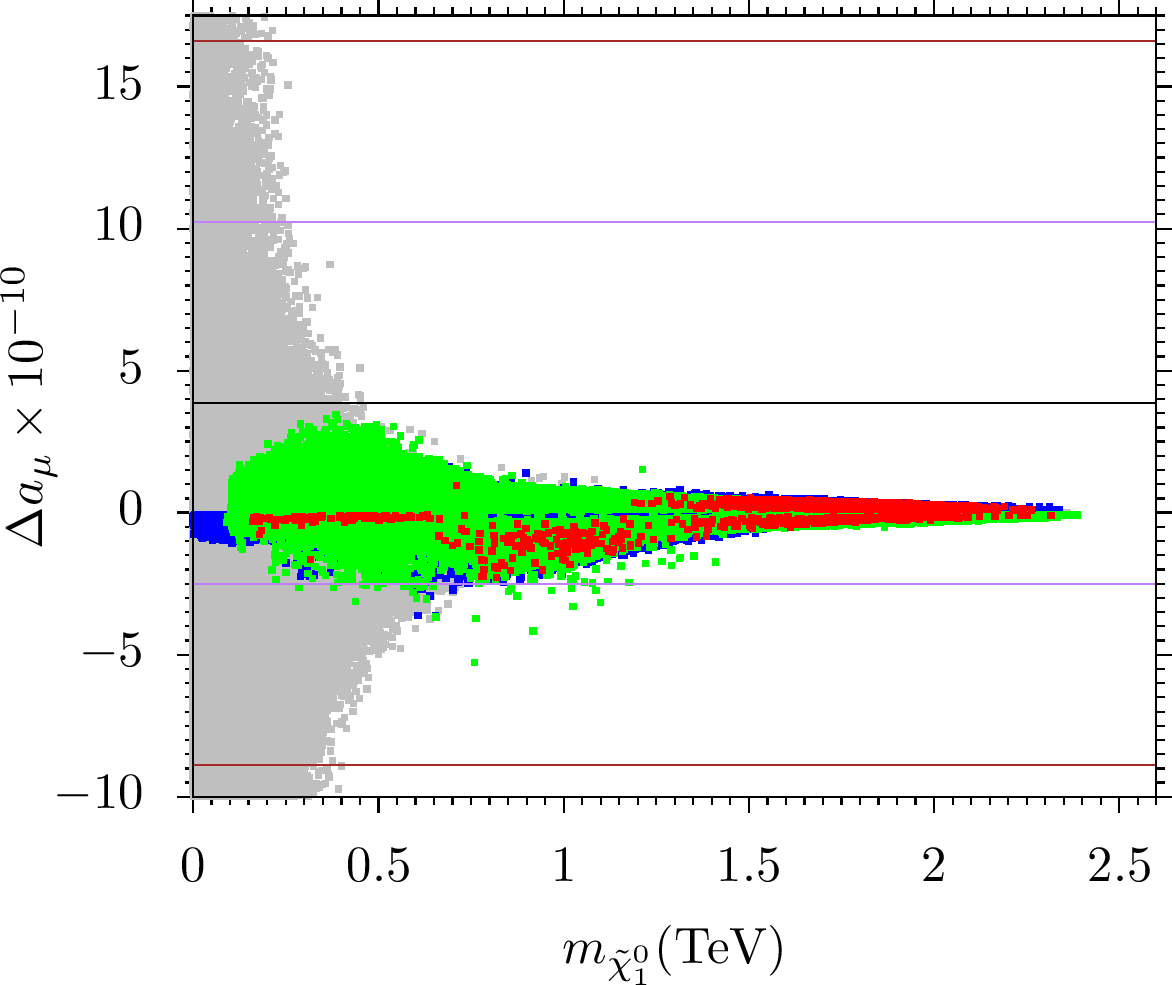}
\caption{\small {Illustration of the lightest neutralino DM candidate vs. the anomalous magnetic moment of the muon, \( a_\mu \). Grey points correspond to parameter sets satisfying REWSB, yielding LSP neutralino. Blue points are a subset of gray points that meet all LHC SUSY
particle mass bounds, B-physics, as well as LHC Higgs mass bounds, indicating over-saturated relic density. Green
points are a subset of blue points that indicate under-saturated relic density. Finally, red points are a subset of
green points and reflect the saturated relic density bound (Planck 2018, 5$\sigma$ bounds). The black curve indicates the experimental central value of \( \Delta a_\mu \), with the purple and brown curves denoting the \(1\sigma\) and \(2\sigma\) confidence regions, respectively.}}
\label{input_params11}
\end{figure}

	\begin{figure}[t]
	\centering \includegraphics[width=7.90cm]{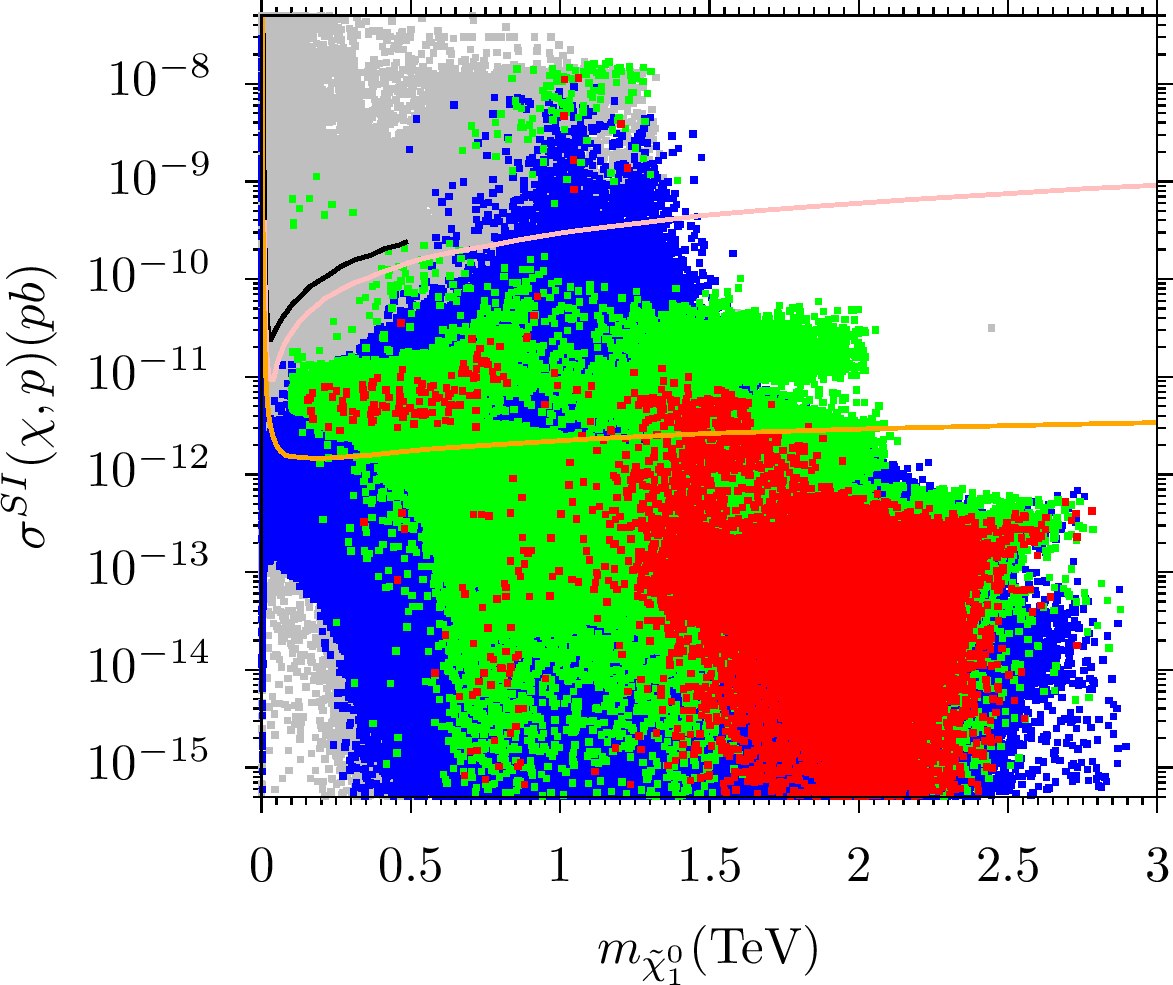}
	\centering \includegraphics[width=7.90cm]{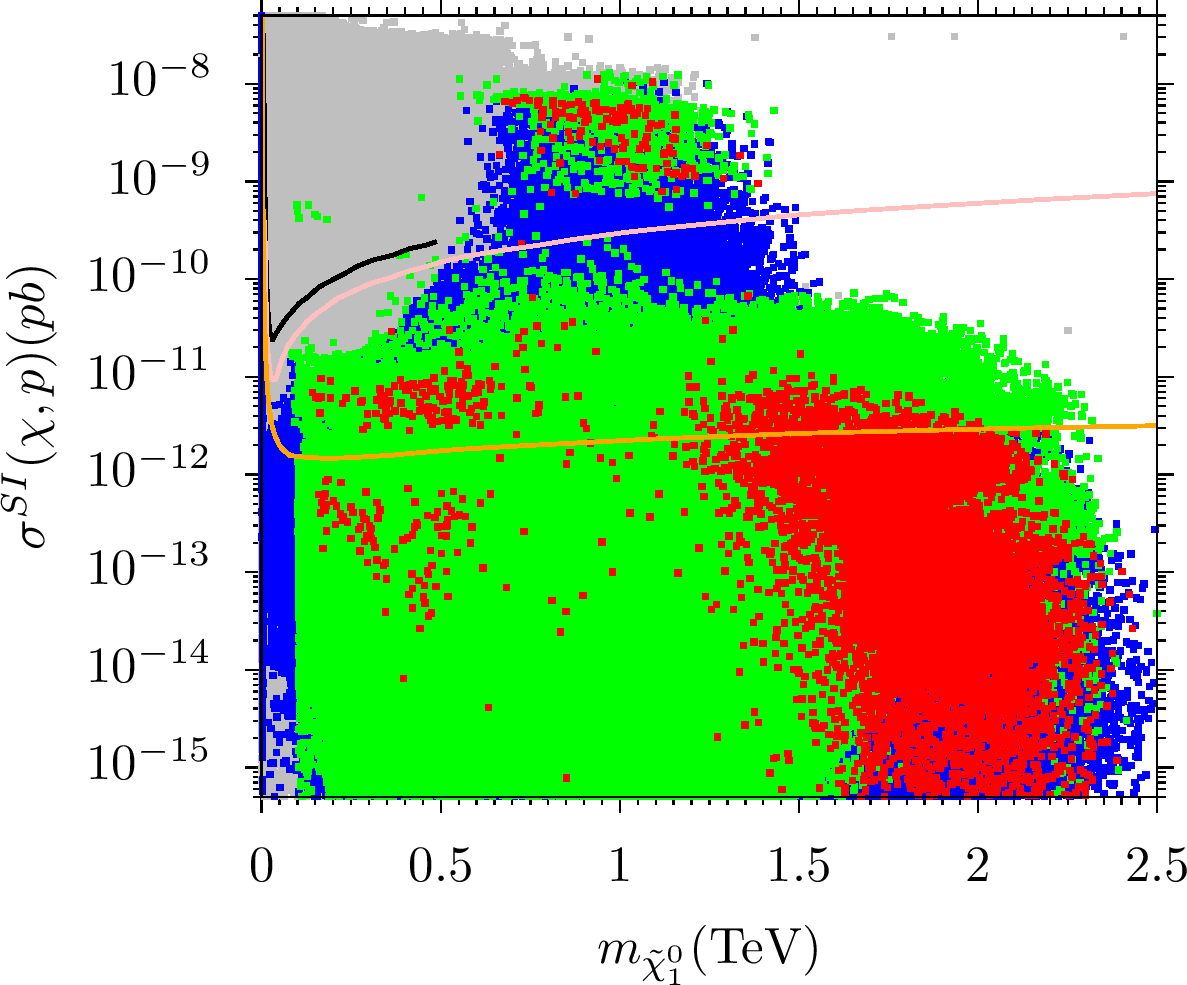}
	\centering \includegraphics[width=7.90cm]{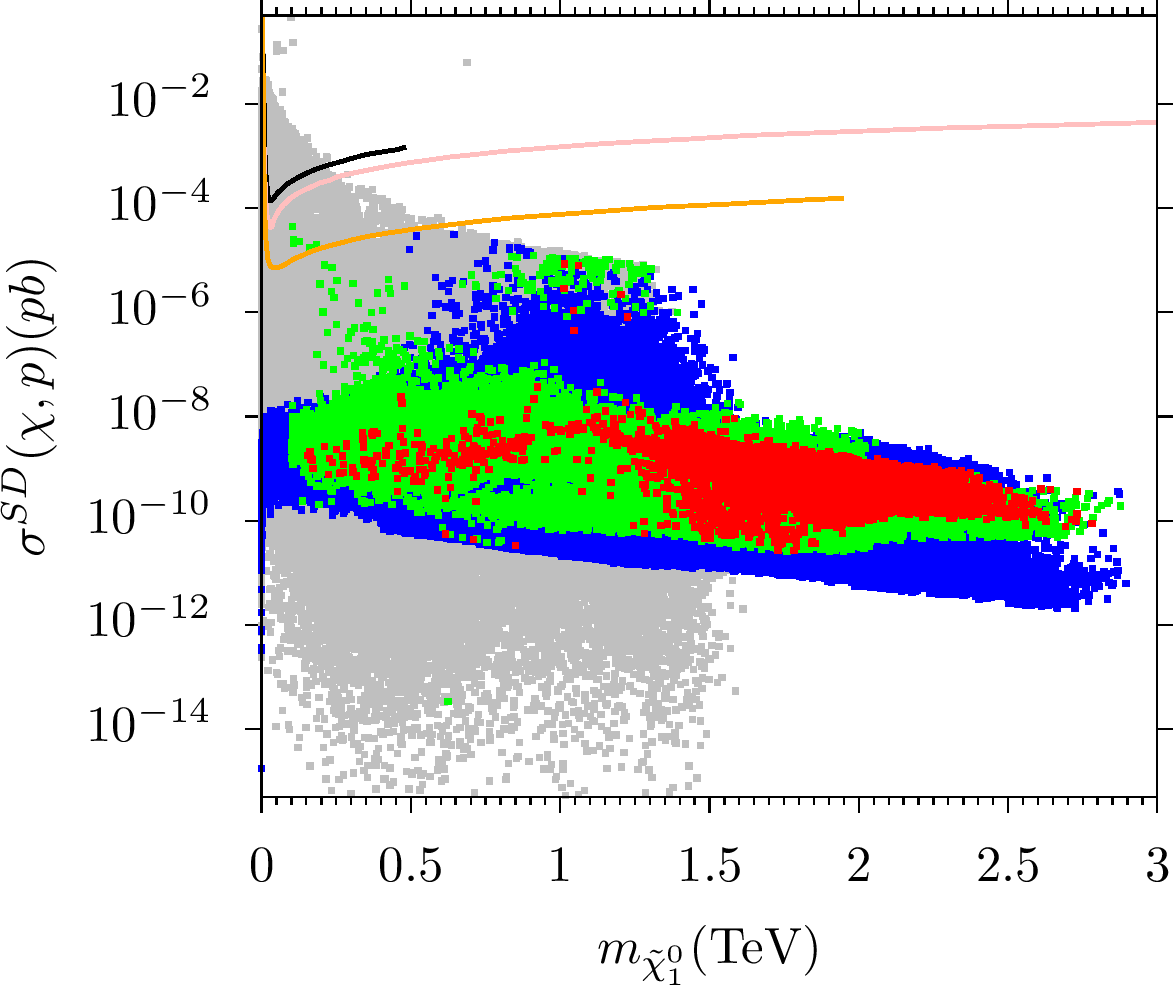}
    \centering \includegraphics[width=7.90cm]{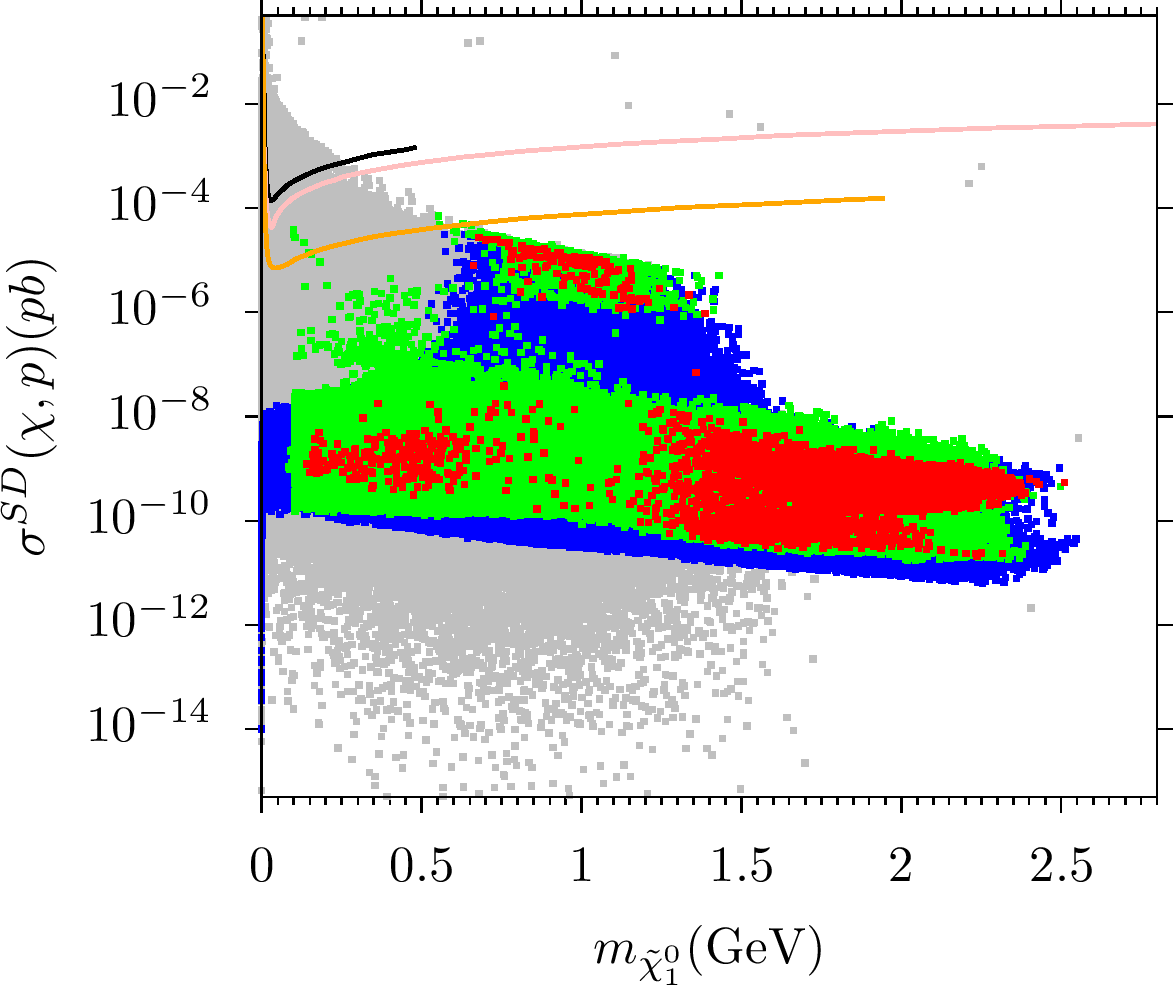}
    \caption{Rescaled spin-independent neutralino-proton scattering cross-section vs. the neutralino mass at the top and spin-dependent neutralino-proton scattering cross-section vs. the neutralino mass at the bottom plots.  In both panels, the solid black lines represent the XENONnT \cite{XENON:2023cxc}, while the pink and orange lines represent the current LZ and LZ-1000 days sensitivity \cite{LZ:2022lsv,LZ:2018qzl}. The left panel is for $\mu <0$ right panel is for $\mu>0$. Grey points complete REWSB and offer LSP neutralino. Blue points are a subset of gray points that meet all LHC SUSY particle mass bounds, B-physics, as well as LHC Higgs mass bounds, indicating an over-saturated relic density.  Green points are a subset of  blue points that indicate under-saturated relic density. Finally, red points are a subset of green points and reflect the saturated relic density bound (Planck 2018, 5$\sigma$ bounds).
		}
		\label{glumu}
\end{figure}
	
\begin{table}[h!]\hspace{-1.0cm}
\centering
\begin{tabular}{|c|cccccccc|}
\hline
\hline
                 & Point 1 & Point 2 & Point 3 & Point 4 & Point 5&Point 6& Point 7& Point 8\\

\hline
$m_{L}$        & 3456.7 & 1899.9   & 1931.8    & 3706.6 & 3938.6&1987.7&2372.8&1779.7\\
$m_{R}$        & 746.87 & 800.88   & 787.77    & 1059.2 & 5986.5 & 2695.9 & 3923.6 & 4152\\
$M_{1} $       & -1640 & -1966.8  & -1989.9   & -1432.4 &4800.5&2520.8&-3292&-2943.9\\
$M_{2}$        & -2767.4 & -2088.9   & -2120.4   & -2794.4 & 2471.8&1621.8&-2188.8&-2256.8\\
$M_{3}$        & -2220.7 & -1776.1   & -1796.9   & -2184.2  &6971.6&3312.7&-4419&-4523.8\\
$A_0$          & 4723.5 & 3980.9  & 4032.3    & 4688 & -6752.9&-4214.5&4905.8&4920.8\\
$\tan\beta$    & 9.31  & 11            & 12      & 16.9 & 36.2&14.5&39.4&39.5\\
$m_{H_u}=m_{H_d}$   & 3507.8  & 3440.7  & 3481    & 3490.3 & 0&2288&0&1122.3\\
\hline
$\mu$            &2576.5   & 929.1        &   999.77        & 2583.8 &8806.8 &4370.6 &5928.8&5797.2\\

\hline

\hline
$m_h$            &126   & 125     & 126       & 127 & 126&124 &125.5&125.4\\
$m_H$            & 4669 & 3542    & 3804   &4450  & 7614&4786&3007&2919\\
$m_A$            & 4638 & 3519    & 3779      &4421  &7565&4755&\textbf{2988}&\textbf{2900}\\
$m_{H^{\pm}}$    & 4670 & 3543    & 3805 &4450  &7615 &4787&3009&2921\\

\hline
$m_{\tilde{\chi}^0_{1,2}}$
                 & \textbf{733}, 2317 & \textbf{809},944    & \textbf{876}, 1015    & \textbf{639}, 2341 &\textbf{2059}, 2197&\textbf{1192},1199&\textbf{1509},1846&\textbf{1388},1694\\

$m_{\tilde{\chi}^0_{3,4}}$
                 & 2585, 2596   & 944, 1643 & 1016, 1763    & 2592, 2606 &8693, 8693&4388,4389&5917,5918&5786,5787\\

$m_{\tilde{\chi}^{\pm}_{1,2}}$
                 & 2315, 2620  & 915, 1520 &  985, 1738&2339,2627    &\textbf{2065}, 8624  &\textbf{1195},4393&1849,5914&1698,5783\\
\hline
$m_{\tilde{g}}$  & 4664 & 3520 &  3810  & 4609 &13629&6913&9020&8888\\
\hline
$m_{ \tilde{u}_{L,R}}$
                 & 5486, 3991 & 3680, 3115 & 3994, 3366  &5606, 3998 & 12112, 12973&6211,6502&8059,8658&7874,8552\\
$m_{\tilde{t}_{1,2}}$
                 & \textbf{768},4699 & \textbf{849}, 2984  & 911, 3237  & 671, 4772 & 10590, 10793&5082,5552&6577,7134&6428,7095\\
\hline $m_{ \tilde{d}_{L,R}}$
                 & 5487, 3977 & 3681, 3095  & 3995, 3344  & 5607, 3987 & 12113, 12945&6212,6483&8059,8638&7875,8536\\
$m_{\tilde{b}_{1,2}}$
                 & 3874, 4763 & 3004, 3024  & 3234, 3283  & 3750, 4843 & 10692, 12354&5522,6383&6642,7476&6496,7416\\
\hline
$m_{\tilde{\nu}_{1,2}}$
                 & 3870 & 2165  & 2369  & 4099 &4279&2174&2811&2559\\
$m_{\tilde{\nu}_{3}}$
                 &  3856 & 2139 & 2336  & 4047 &3787&2118&2507&2258\\
\hline
$m_{ \tilde{e}_{L,R}}$
                & 3866, 899 & 2166, 985 &2370, 1064  & 4094, 1138 & 4279, 6231&2182,2882&2804,4170&2550,4133  \\
$m_{\tilde{\tau}_{1,2}}$
                & 776, 3854 & 868, 2141&  \textbf{910}, 2338 & \textbf{662},4044  &3777, 5494&2124,2798&2503,3730&2254,3732 \\
\hline

$\sigma_{SI}({\rm pb})$
                & $2.6\times 10^{-13}$ & $ 7.7\times 10^{-10} $ & $ 7.5\times 10^{-10}$ & $6.3\times 10^{-13} $ &$4.9\times 10^{-14} $ &$1.0\times 10^{-11} $&$3.1\times 10^{-13} $&$2.9\times 10^{-13} $\\

$\sigma_{SD}({\rm pb})$
                & $1.5\times 10^{-9}$ & $2.4\times 10^{-6}$ & $2\times 10^{-6}$ & $1.5\times 10^{-9} $ &
$1.8\times 10^{-10} $&$6.3\times 10^{-9} $&$6.9\times 10^{-11} $&$7.6\times 10^{-11} $\\

$\Delta a_\mu$ & $-1.1\times 10^{-11}$ & $-4.4\times 10^{-11} $ 
                & $-4\times 10^{-12}$ & $-1.9\times 10^{-11}$ &$2.2\times 10^{-11} $
&$3.5\times 10^{-11} $&$-5.3\times 10^{-11} $
&$-6\times 10^{-11} $\\

$\Omega_{CDM}h^{2}$&  0.123 & 0.123   & 0.119  & 0.117 & 0.118&0.118&0.117&0.117\\
\hline
\hline
\end{tabular}
\caption{All masses are given in units of GeV, and the analysis corresponds to the $\mu>0$ scenario. All points satisfy the LHC SUSY sparticle mass bounds, B-physics constraints, the LHC Higgs mass bound, and the $5\sigma$ relic density limit from Planck 2018, as detailed in Section~\ref{constraints}. Points 1 $\&$ 2 represent neutralino-stop coannihilation, points 3 $\&$ 4 correspond to neutralino-stau coannihilation, and points 5 $\&$ 6 represent neutralino-chargino coannihilation.  Finally, points 7 $\&$ 8 correspond to Higgs resonance.
}
\label{table1}
\end{table}

\begin{table}[h!]\hspace{-1.0cm}
\centering
\begin{tabular}{|c|cccccc|}
\hline
\hline
                 & Point 1 & Point 2 & Point 3 & Point 4 & Point 5& Point 6\\

\hline
$m_{L}$        & 3610 & 2290   & 2419.3    & 0 & 3969.3&4302.4\\
$m_{R}$        & 1014.5 & 900.2   & 742.27    & 7572.1 & 5819.4&6299\\
$M_{1} $       & -2186 & -2098.9  & -2978.9   & 1920.7 &5587.4&5739.9\\
$M_{2}$        & -2994.2 & -2318   & -2380.3   & 3031.3 & 3018.9&3100\\
$M_{3}$        & -2542.9 & -2013.4   & -2027.6   & 6213.8  &6957.8&7477.1\\
$A_0$          & 5436.1 & 4398.9  & 4450.5    & -6387.2 & -7909.4&-8130.3\\
$\tan\beta$    & 7.2  & 8.8            & 8.93     & 28.7 & 19.8&19.2\\
$m_{H_u}=m_{H_d}$   & 4293.2  & 3699.2  & 3697    & 5533.8 & 1832.9&596.9\\
\hline
$\mu$            &-2585.7   & -1270.1        &   -1589.8        & -6879 & -8879&-9564\\
\hline

\hline
$m_h$            &126   & 125.5     & 125.8       & 125.6 & 125.5&125.7\\
$m_H$            & 5399 & 4000    & 4295      &7810  &8763&9299\\
$m_A$            & 5364 & 4073    & 4267    &7759  &8706&9239\\
$m_{H^{\pm}}$    & 5399 & 4100    & 4296     &7810  &8764&9200\\

\hline
$m_{\tilde{\chi}^0_{1,2}}$
                 & \textbf{982}, 2472 & \textbf{913}, 1276    & \textbf{924}, 1588    & \textbf{851}, 2509 & \textbf{2529},2573&\textbf{2598},2648\\

$m_{\tilde{\chi}^0_{3,4}}$
                 & 2587, 2629   & 1281, 1866 & 1597, 1990  & 6796, 6796 &8772,9450&8773,9550 \\

$m_{\tilde{\chi}^{\pm}_{1,2}}$
                 & 2466, 2657  & 1301, 1845 &  1619, 1669    &2512, 6725  &\textbf{2535},8714&\textbf{2604},9384\\
\hline
$m_{\tilde{g}}$  & 5289 & 4266 &  4268  & 12258 &13581&14539\\
\hline
$m_{ \tilde{u}_{L,R}}$
                 & 6008, 4576 & 4266, 3557 & 4591, 3720  & 10339, 12821 & 12132,12872&12977,13792\\
$m_{\tilde{t}_{1,2}}$
                 & \textbf{1022}, 5083 & \textbf{956}, 3489  & 954, 3797  & 8457, 10541 & 10342,10800&11176,11603\\
\hline $m_{ \tilde{d}_{L,R}}$
                 & 6009, 4556 & 3533, 4267  & 4592,3696 &10339,12830 & 12132,12830&12978,13751\\
$m_{\tilde{b}_{1,2}}$
                 & 4466, 5164 & 3451, 3543  & 3608, 3855  & 8608, 12244 & 10765,12498&11568,13424\\
\hline
$m_{\tilde{\nu}_{1,2}}$
                 & 4075 & 2571  & 2861  & 1881 &4465&4784\\
$m_{\tilde{\nu}_{3}}$
                 &  4065 & 2552 & 2842  & 894 &4200&4628\\
\hline
$m_{ \tilde{e}_{L,R}}$
                & 4072, 1252 & 2572, 1103 &2860, 1044  & 1756, 7599 & 4471,6159&4789,6629  \\
$m_{\tilde{\tau}_{1,2}}$
                & 1186, 4064 & 1017,2554 &  \textbf{944}, 2843 & \textbf{865}, 7221 &4310,5911&4639,6396 \\
\hline

$\sigma_{SI}({\rm pb})$
                & $1.1\times 10^{-11}$ & $ 2.6\times 10^{-10} $ & $ 6.6\times 10^{-11}$ & $1.3\times 10^{-14} $ & 
 $1.5\times 10^{-13} $& $6.5\times 10^{-14} $\\

$\sigma_{SD}({\rm pb})$
                & $2.1\times 10^{-9}$ & $1.9\times 10^{-7}$ & $3.7\times 10^{-8}$ & $3.3\times 10^{-11} $ &
$-2.1\times 10^{-10} $&$-1.4\times 10^{-10} $\\
$\Delta a_\mu$
                & $7.1\times 10^{-12}$ & $2.4\times 10^{-11}$ & $1.8\times 10^{-11}$ & $-3.2\times 10^{-11} $ &$-1.08\times 10^{-11} $
&$-9.2\times 10^{-11} $\\

$\Omega_{CDM}h^{2}$&  0.125 & 0.125   & 0.117  & 0.118 & 0.119&0.114\\
\hline
\hline
\end{tabular}
\caption{All masses are given in units of GeV, and the analysis corresponds to the $\mu<0$ scenario. All points satisfy the LHC SUSY sparticle mass bounds, B-physics constraints, the LHC Higgs mass bound, and the $5\sigma$ relic density limit from Planck 2018, as detailed in Section~\ref{constraints}. Points 1 $\&$ 2 represent neutralino-stop coannihilation, points 3 $\&$ 4 correspond to neutralino-stau coannihilation, and points 5 $\&$ 6 represent neutralino-chargino coannihilation.}

\label{table2}
\end{table}

 \end{widetext}
     
 
 In the $\Theta_1$--$\Theta_2$ plane of Fig.~\ref{input_params1}, for the $\mu<0$ scenario (left panel), the red points are predominantly localized within the range $-0.8 \lesssim \Theta_1 \lesssim 0.8$, with a noticeable clustering along the off-diagonal regions. For $\Theta_2$, the red points are concentrated mainly in the intervals $0.4 \lesssim \Theta_2 \lesssim 0.7$ on the positive axis, and a slightly less dense clustering is observed on the negative side within $-0.9 \lesssim \Theta_2 \lesssim -0.3$. Green points are distributed more uniformly across the plane, although their density also increases near the off-diagonal regions, similar to the distribution of red points. A comparable distribution pattern is observed for the $\mu>0$ scenario (right panel).

In the $\Theta_1$--$\Theta_3$ plane, the distributions of red and blue points largely mirror the patterns seen in the $\Theta_1$--$\Theta_2$ plane, although fewer points are concentrated along the diagonal for both $\mu<0$ and $\mu>0$ cases. In contrast to the $\Theta_1$--$\Theta_2$ plane, where the blue points display a dipole-like distribution, the blue points in the $\Theta_1$--$\Theta_3$ plane are more uniformly dispersed.

Finally, in the $\Theta_2$--$\Theta_3$ plane, for $\mu<0$, the distribution closely resembles that observed in the $\Theta_1$--$\Theta_2$ plane. However, for the $\mu>0$ case, the points exhibit a clear dipole-like structure along the off-diagonal, with an absence of solutions near the origin.

We numerically analyze the SUSY breaking solutions by solving the complete set of equations presented in Sections~2 and~3. The color scheme and panel structure in Fig.~\ref{funda_params2} are consistent with those used in Fig.~\ref{input_params1}. Given the pivotal role of gaugino masses in renormalization group equations (RGEs), their behavior is illustrated in Fig.~\ref{funda_params2}. The upper left and right panels correspond to the $\mu < 0$ and $\mu > 0$ scenarios, respectively.

In the $M_1$--$M_2$ plane, red points span a broad range of $M_1$ values from $-4$ to $6$~TeV, with a higher concentration observed in the intervals $[2,6]$~TeV and $[-4,-2]$~TeV for both $\mu < 0$ and $\mu > 0$. Notably, the $\mu < 0$ scenario exhibits a slightly extended overall range. For $M_2$, solutions—especially the red points—tend to cluster around $[1,4]$~TeV and $[-4,-1]$~TeV, again similarly distributed for both signs of $\mu$. Comparable patterns are observed in the $M_1$--$M_3$ and $M_2$--$M_3$ planes, with a modest extension into the positive range, particularly evident in the $\mu < 0$ case.

Figure~\ref{funda_params3} presents the parameter correlations in the $m_L$--$m_R$, $\tan\beta$--$m_{H_{u,d}}$, and $m_{\tilde{\chi}_1^0}$--$m_{\tilde{\chi}_1^{\pm}}$ planes, where the left and right panels correspond to $\mu < 0$ and $\mu > 0$, respectively, following the same color coding as in Fig.~\ref{input_params1}.

In the $m_L$--$m_R$ plane, red points exhibit clear clustering behavior. For $\mu < 0$, the $m_L$ values predominantly fall within $[1,5.5]$~TeV, while for $\mu > 0$, the clustering appears slightly narrower, spanning $[1.5,4.5]$~TeV. The $m_R$ distribution shows dense clustering in the range $[0.5,7]$~TeV for $\mu < 0$, and a similar but slightly more constrained spread of $[0.5,6]$~TeV for $\mu > 0$. These regions indicate the favored parameter space for left- and right-handed scalar mass parameters.

In the $\tan\beta$--$m_{H_{u,d}}$ plane, a higher density of red points is observed for the $\mu < 0$ scenario, particularly within $10 \lesssim \tan\beta \lesssim 40$ and $0 \lesssim m_{H_{u,d}} \lesssim 5.5$~TeV. The $\mu < 0$ case shows an extended $m_{H_{u,d}}$ spectrum compared to $\mu > 0$. While the preferred $\tan\beta$ range for $\mu < 0$ lies within $10 \lesssim \tan\beta \lesssim 40$, the $\mu > 0$ case allows for values up to $60$, i.e., $10 \lesssim \tan\beta \lesssim 60$.


In our parameter scans, we identify a region characterized by chargino-neutralino coannihilation, depicted in the lower panel of Fig.~\ref{funda_params3}, where multiple experimental constraints are presented in the $m_{\tilde{\chi}_1^{\pm}}$--$m_{\tilde{\chi}_1^0}$ plane. The red points indicate scenarios in which the lightest chargino are closely mass-degenerate with the LSP neutralino, with the chargino mass falling within the range $0.6~\mathrm{TeV} \lesssim m_{\tilde{\chi}_1^{\pm}} \lesssim 2.8~\mathrm{TeV}$. Our results align well with those reported in Ref.~\cite{Gomez:2020gav}. Additionally, based on recent experimental findings from the ATLAS Collaboration~\cite{ATLAS:2021ilc}, which investigate slepton- and Standard Model boson-mediated decays of $\tilde{\chi}_1^{+}\tilde{\chi}_1^{-}$ and $\tilde{\chi}_1^{\pm}\tilde{\chi}_2^0$ systems, scenarios with charginos nearly degenerate with the LSP and masses exceeding 300\,GeV remain consistent with Current excluding limitations at a 95\% level of confidence. It is essential to highlight that in regions of parameter space where sleptons appear heavier than the charginos, slepton-mediated decay channels are kinematically suppressed. Given that our scans also yield viable solutions with heavier NLSP charginos, we anticipate that future LHC searches will have the potential to explore these regions further.

The top panel of Fig.~\ref{delew} illustrates the stau-neutralino coannihilation region, with the left and right panels corresponding to the scenarios $\mu < 0$ and $\mu > 0$, respectively. Our analysis reveals that the lightest stau, almost degenerate in mass with the lightest neutralino, spans a mass range of approximately $0.2~\mathrm{TeV} \lesssim m_{\tilde{\tau}_1} \lesssim 2.4~\mathrm{TeV}$. These findings are in good agreement with the results presented in Refs.~\cite{Raza:2018jnh,Gomez:2020gav}. Furthermore, our viable parameter space is consistent with the constraints from the CMS search based on $137~\mathrm{fb}^{-1}$ of data at $\sqrt{s} = 13~\mathrm{TeV}$~\cite{CMS:2022rqk}. We anticipate that forthcoming data from Run 3 of the LHC and future collider experiments will further probe the parameter space explored in this study.

In the middle panel of Fig.~\ref{delew}, we present the distribution of data points in the $m_{\tilde{t}_1}$--$m_{\tilde{\chi}_1^0}$ plane. The color scheme employed is consistent with that of Fig.~\ref{input_params1}. From our current parameter scans, we observe that NLSP stop solutions extend across the mass range $0.2\,\mathrm{TeV} \leq m_{\tilde{t}_1} \leq 1.2\,\mathrm{TeV}$. It is worth highlighting that a previous analysis~\cite{Khan:2023yjs} identified such solutions only up to approximately 0.9\,TeV. In our findings, particularly for the red-colored viable points, the mass gap between an NLSP stop with the LSP neutralino may reach values as high as 300\,GeV. However, this occurs in scenarios where both stop and neutralino masses are large, thereby ensuring that the relative mass difference $\Delta m_{{\rm NLSP, LSP}}/m_{\rm LSP}$ remains within the 10\% threshold. A substantial mass gap enables two-body decay modes such as $\tilde{t}_1 \rightarrow t\,\tilde{\chi}_1^0$, in addition to the three-body decay $\tilde{t}_1 \rightarrow W\,b\,\tilde{\chi}_1^0$ as well as four-body decay $\tilde{t}_1 \rightarrow f\,f'\,b\,\tilde{\chi}_1^0$. Conversely, in the compressed mass spectrum regime, where these decay channels are kinematically inaccessible, the loop-induced two-body decay $\tilde{t}_1 \rightarrow c\,\tilde{\chi}_1^0$ typically turns into the dominating mode, as discussed in Refs.~\cite{Hikasa:1987db,Muhlleitner:2011ww}. Furthermore, even in scenarios with a small mass splitting that permits $\tilde{t}_1 \rightarrow t\,\tilde{\chi}_1^0$, stop masses up to 550\,GeV have already been ruled out by current LHC analyses~\cite{ATLAS:2021kxv}. The NLSP stop candidates presented in this work lie beyond these existing exclusion limits, thereby remaining viable under current experimental constraints. We anticipate that future LHC searches will further probe these parameter regions.

In addition to the coannihilation mechanisms, we also identify scenarios in which the lightest supersymmetric particle (LSP), specifically the neutralino\(\tilde{\chi}_1^0\), achieves the observed relic density through resonant annihilation via heavy Higgs bosons. In such cases, a pair of neutralinos can annihilate via \(s\)-channel exchange of the CP-odd Higgs boson \(A\), or the CP-even Higgs bosons \(H\) and \(h\), subsequently decay into Standard Model final states. These so-called Higgs funnel or resonance solutions become significant when the mass condition \(m_{A} \approx 2m_{\tilde{\chi}_1^0}\) is satisfied. As shown in the third panel of the Figure~\ref{delew} illustrates that such a mass alignment is realized in our scans, particularly for the case of \(\mu > 0\), where the mass relation \(m_A \approx m_H\) also holds, consistent with expectations from the decoupling regime of the MSSM Higgs sector. Current experimental searches place strong constraints on this scenario. As reported in Ref.~\cite{CMS:2022goy}, the decay channel \(A, H \rightarrow \tau\bar{\tau}\) excludes pseudoscalar Higgs masses below approximately 1.7~TeV for \(\tan\beta \lesssim 30\). Furthermore, projected sensitivity at the LHC Run 2, Run 3, as well as High-Luminosity LHC (HL-LHC) show that for the \(\tan\beta \lesssim 10\), values of \(m_A\) near 1~TeV, 1.1~TeV, and 1.4~TeV, respectively, can be probed~\cite{Baer:2022qqr, Baer:2022smj}. Our analysis reveals that the viable \(A\)-resonance solutions predominantly lie in the mass interval of 1.2~TeV to 1.7~TeV for the \(\mu > 0\) case. Notably, such solutions are absent in the \(\mu < 0\) scenario, underlining the crucial dependence of this annihilation channel on the sign of the higgsino masses parameter. Given the present and anticipated experimental sensitivity, the parameter space corresponding to this channel is already under significant pressure, with most of it either excluded or within the reach of forthcoming LHC runs.
{
The anomalous magnetic moment of the muon, \( a_\mu \equiv (g-2)_\mu/2 \), has long exhibited a discrepancy between its Standard Model (SM) prediction and experimental measurements. The 2020 White Paper (WP20) reported the SM value as \cite{Aoyama:2020ynm}
\[
a_\mu^{\text{SM, WP20}} = 116\,591\,810(43) \times 10^{-11},
\]
derived predominantly from data-driven evaluations of the leading-order hadronic vacuum polarization (LO HVP) using \( e^+e^- \) annihilation data. The Muon \( (g{-}2) \) Experiment at Fermilab (E989) has recently released a new high-precision determination of the anomalous magnetic moment of the positive muon, \( a_\mu \), utilizing data collected between 2020 and 2023~\cite{Muong-2:2023cdq, Muong-2:2021ojo, Muong-2:2025xyk}:
\[
a_\mu = 116\,592\,0710(162) \times 10^{-12} \quad \text{(139 ppb)}.
\]
Incorporating earlier data, the combined result refines the estimate to~\cite{Muong-2:2025xyk}:
\[
a_\mu = 116\,592\,0705(148) \times 10^{-12} \quad \text{(127 ppb)}.
\]
The updated world average, predominantly driven by Fermilab's measurements, is reported as~\cite{Muong-2:2025xyk}:
\[
a_\mu^{\text{exp}} = 116\,592\,0715(145) \times 10^{-12} \quad \text{(124 ppb)},
\]
marking a more than fourfold enhancement in experimental precision relative to previous evaluations.
Concurrently, substantial progress in lattice QCD calculations has enabled a more accurate theoretical prediction of the leading-order hadronic vacuum polarization (LO HVP) contribution. A recent consensus estimate now achieves a precision of approximately 0.9\%, leading to an updated Standard Model (SM) value for the muon anomaly~\cite{Aliberti:2025beg}:
\[
a_\mu^{\text{SM}} = 116\,592\,033(62) \times 10^{-11} \quad \text{(530 ppb)}.
\]
The resulting discrepancy between the experimental average and the SM prediction is found to be~\cite{Muong-2:2025xyk, Aliberti:2025beg}:
\[
\Delta a_\mu = a_\mu^{\text{exp}} - a_\mu^{\text{SM}} = 38.5(63.7) \times 10^{-11},
\]
corresponding to a statistical deviation of only \( 0.6\sigma \). This marks a substantial reduction from the previously reported \( 4.2\sigma \) discrepancy and suggests that the experimental measurements are now in good agreement with SM predictions at the current level of precision.
The major shift originates from replacing the earlier \( e^+e^- \)-based LO HVP estimate with a more precise lattice QCD determination, thereby resolving the earlier discrepancy and bringing the SM prediction into alignment with experimental observations. In the context of SUSY GUTs, the sign of the Higgsino mass parameter \( \mu \) is one of the fundamental parameters regarding the phenomenological studies. Figure~\ref{input_params11} presents results from a random scan of intersecting D6-brane models, compactified on the $\mathbf{T^6/(\mathbb{Z}_2 \times \mathbb{Z}_2)}$ orientifold.
Our analysis indicates that supersymmetric scenarios with a negative Higgsino mass parameter, \( \mu < 0 \), yield contributions to \( (g-2)_\mu \) consistent with the experimental measurement within the \( 1\sigma \) range for lightest neutralino masses up to 500~GeV. In contrast, for \( \mu > 0 \), the \( 1\sigma \) compatibility extends to neutralino masses as high as 1~TeV.
}

In Fig.~\ref{glumu}, we show the spin-independent (SI) and spin-dependent (SD) neutralino-proton scattering cross sections as functions of the lightest neutralino mass for the cases of \(\mu < 0\) (left panel) and \(\mu > 0\) (right panel). The choice of the sign of the higgsino mass parameter \(\mu\) significantly influences the neutralino composition, and consequently, its interaction strength with Standard Model particles. In particular, the sign of \(\mu\) plays a critical role in determining the relative higgsino-gaugino mixing, which directly impacts the scattering cross section relevant to DM direct detection. In both panels of Fig.~\ref{glumu}, we include current and projected exclusion limits from leading direct detection experiments. The solid blue curve represents the sensitivity of the XENONnT experiment~\cite{XENON:2023cxc}, the pink line corresponds to the 2022 LUX-ZEPLIN (LZ) results~\cite{LZ:2022lsv}, and the black line displays the anticipated reach of the 1000-day exposure of the LZ experiment~\cite{LZ:2018qzl}. From the plots in the \(m_{\tilde{\chi}^0_1} - \sigma_{\text{SI}}\) plane, we observe that nearly all solutions consistent with the Planck 2018 DM relic density bounds lie below the current exclusion limits set by XENONnT and LZ. However, a small subset of points associated with the chargino-neutralino coannihilation scenario approach or slightly exceed these bounds, highlighting regions of interest for near-future searches.

Similarly, the plots in the \(m_{\tilde{\chi}^0_1} - \sigma_{\text{SD}}\) plane reveal that our solutions are compatible with current experimental constraints and remain well within the discovery reach of upcoming direct detection experiments. The continued improvements in experimental sensitivity, particularly from long-duration exposures such as the projected 1000-day LZ dataset, are expected to probe significant portions of the current permitted parameter space, especially in scenarios where the lightest neutralino exhibits mixed higgsino-gaugino characteristics. It is noteworthy that the $\mu < 0$ scenario exhibits a comparatively richer parameter space in certain regions. For instance, the mass of the lightest neutralino, $m_{\tilde{\chi}_1^0}$, extends from approximately 2.5~TeV to 2.9~TeV in this case. The same extension has been seen for the whole spectrum for the $\mu < 0$ scenario. However, despite this broader mass range, only a limited number of parameter points for $\mu < 0$ have been probed by current DD DM experiments, particularly in scenarios involving chargino-neutralino coannihilation. In contrast, the $\mu > 0$ case shows a more extensively explored parameter space under similar conditions, with significantly more points falling within the reach of existing DD constraints.

Finally, we present two tables of benchmark points. Table~\ref{table1} summarizes representative solutions illustrating coannihilation scenarios and the resonance annihilation mechanism for the $\mu < 0$ case. Points 1 and 2 correspond to examples of NLSP stop coannihilation. In these points, the NLSP stop masses are approximately $0.77$~TeV and $0.85$~TeV, respectively, while the LSP neutralino, predominantly bino-like with a small higgsino admixture, has masses of about $0.73$~TeV and $0.8$~TeV. Points 3 and 4 illustrate the NLSP stau coannihilation scenario. Here, the stau masses are around $0.91$~TeV and $0.66$~TeV, with the corresponding LSP neutralino masses being $0.87$~TeV and $0.64$~TeV, respectively. The LSP is mostly bino-like in Point 3, while in Point 4 it contains a significant higgsino component. Points 5 and 6 depict chargino-neutralino coannihilation. In these cases, the lightest chargino masses are $m_{\tilde{\chi}_1^\pm} = 2.056$~TeV and $1.195$~TeV, closely matched with the LSP neutralino masses of $2.059$~TeV and $1.192$~TeV, respectively. The LSPs in these points are predominantly bino-like with a notable wino admixture. Finally, Points 7 and 8 represent examples of $A$-resonance solutions. The LSP neutralino masses are approximately $1.509$~TeV and $1.388$~TeV, while the corresponding pseudoscalar Higgs masses are $m_A = 2.988$~TeV and $2.900$~TeV, respectively.

Table~\ref{table2} presents representative benchmark points corresponding to coannihilation and resonance annihilation scenarios for the $\mu > 0$ case. Points 1 and 2 illustrate the NLSP stop coannihilation mechanism. In these cases, the NLSP stop masses are approximately $1.022$~TeV and $0.956$~TeV, respectively, while the LSP neutralino, primarily bino-like, has masses around $0.98$~TeV and $0.91$~TeV. Points 3 and 4 correspond to NLSP stau coannihilation. The stau masses are around $0.944$~TeV and $0.865$~TeV, while the LSP neutralino, which is a bino with a masses, is approximately $0.924$~TeV and $0.852$~TeV, respectively. Points 5 and 6 represent examples of chargino-neutralino coannihilation. Here, the lightest chargino masses are $m_{\tilde{\chi}_1^\pm} = 2.535$~TeV and $2.603$~TeV, closely matched with the LSP neutralino masses of $2.529$~TeV and $2.598$~TeV, respectively. In these scenarios, the neutralino is predominantly wino-like with a subdominant bino component.

\section{Conclusion}
{
Because of the given emerging consistency between the experimental determination of the muon's anomalous magnetic moment and the SM prediction, we perform a systematic exploration of the model’s parameter space for both signs of the Higgsino mass parameter, \( \mu > 0 \) and \( \mu < 0 \). The updated comparison between theoretical and experimental values has effectively resolved the previously reported 4.2\( \sigma \) discrepancy, indicating no significant deviation from the SM at the current level of precision. In this context, supersymmetric scenarios with a negative Higgsino mass parameter, \( \mu < 0 \), have gained renewed phenomenological relevance.}
This comprehensive analysis investigates the phenomenological implications of SUSY within a globally consistent framework inspired by string theory, particularly focusing on intersecting D-brane constructions according to the Pati-Salam gauge symmetry. The model is used to develop in Type IIA string theory compactified on a $\mathbf{T^6/(\mathbb{Z}_2 \times \mathbb{Z}_2)}$ orientifold with intersecting D6-branes. Incorporating soft SUSY-breaking terms derived from the $F$-terms of the dilaton and complex structure moduli, the study systematically explores the model’s parameter space, taking into account constraints from collider data, cosmological observations, and direct detection experiments. {
Given the potential consistency of the muon's anomalous magnetic moment with the SM prediction, we perform a thorough exploration of the model's parameter space for both signs of the Higgsino mass parameter, \( \mu < 0 \) and \( \mu > 0 \).} {
Our analysis indicates that supersymmetric scenarios with a negative Higgsino mass parameter, \( \mu < 0 \), yield contributions to \( (g-2)_\mu \) consistent with the experimental measurement within the \( 1\sigma \) range for lightest neutralino masses up to 500~GeV. In contrast, for \( \mu > 0 \), the \( 1\sigma \) compatibility extends to neutralino masses as high as 1~TeV.
} In the $\mu > 0$ case, the neutralino relic abundance is primarily governed by resonant annihilation via heavy Higgs bosons with $m_{A/H} \approx 2$~TeV, and coannihilation processes involving either nearly mass-degenerate charginos ($m_{\tilde{\chi}_1^\pm} \sim 0.7$--$2.5$~TeV) or staus ($m_{\tilde{\tau}_1} \sim 0.2$--$1.8$~TeV). In contrast, the $\mu < 0$ scenario excludes Higgs funnel annihilation mechanisms but permits extended chargino-neutralino coannihilation up to $m_{\tilde{\chi}_1^\pm} \lesssim 2.8$~TeV, with viable stau coannihilation solutions extending to $m_{\tilde{\tau}_1} \lesssim 2.5$~TeV. The resulting sparticle spectrum features gluinos with masses up to 18~TeV and first-generation squarks reaching 16~TeV, while light sleptons remain below 6~TeV. Notably, third-generation sparticles such as stops ($m_{\tilde{t}_1} \sim 0.15$--$1.2$~TeV) lie in compressed mass regions, allowing them to evade current LHC exclusion limits. 

Despite current experimental exclusions ruling out broad regions of the parameter space, substantial viable domains persist, particularly in coannihilation regimes and heavy Higgs resonance scenarios. Anticipated advancements from upcoming experimental efforts—including the High-Luminosity LHC (HL-LHC) as well as next-generation DM detectors like LUX-ZEPLIN with 1000-day exposure—are expected to critically test the remaining parameter space. Overall, this work enhances the phenomenological credibility of string-motivated supersymmetric constructions, offering concrete predictions and experimentally accessible benchmarks. It contributes a coherent strategy for probing TeV-scale new physics, reinforcing the synergy between ultraviolet-complete theories and precision phenomenological studies.

	\section*{Acknowledgements}
	I.K. acknowledges support from Zhejiang Normal University through a postdoctoral fellowship under Grant No.~YS304224924.
    TL is supported in part by the National Key Research and Development Program of China Grant No. 2020YFC2201504, by the Projects No. 11875062, No. 11947302, No. 12047503, and No. 12275333 supported by the National Natural Science Foundation of China, by the Key Research Program of the Chinese Academy of Sciences, Grant No. XDPB15, by the Scientific Instrument Developing Project of the Chinese Academy of Sciences, Grant No. YJKYYQ20190049, and by the International Partnership Program of Chinese Academy of Sciences for Grand Challenges, Grant No. 112311KYSB20210012. 
	

\end{document}